%% file: main_v2.tex
\documentclass[
  notitlepage,
  twocolumn,
  preprintnumbers,
  longbibliography,
  superscriptaddress,
  aps,
  prl,
  amsmath,
  amssymb,
]{revtex4-2}

\usepackage{booktabs} 
\usepackage{physics}
\usepackage{tikz}
\usepackage{tkz-euclide}
\usetikzlibrary{positioning}
\usetikzlibrary{decorations.markings}
\usepackage{ifthen}
\usepackage{braket}

\usepackage{amsmath}
\usepackage{amsthm, amssymb}
\usepackage{slashed} 
\usepackage{graphicx}
\usepackage{xcolor}
\usepackage{bm}
\usepackage[
  pdfpagelabels,
  plainpages=false,
  bookmarks=true,
  colorlinks=true,
  linkcolor=red,
  urlcolor=blue,
  citecolor=blue
]{hyperref}
\usepackage{dsfont}
\usepackage{changepage}
\usepackage{array}
\usepackage{bbm}
\renewcommand{\vec}[1]{\bm{\mathbf{#1}}}
\newcommand{\Z}{\mathbb{Z}}

\theoremstyle{definition}

\theoremstyle{remark}

\newtheorem{fact}{Fact}
\newcommand{\change}[1]{#1}
\begin{document}

\title{Accurate Gauge-Invariant Tensor Network Simulations for Abelian Lattice Gauge Theory in (2+1)D\change{: ground state and real-time dynamics}}

\author{Yantao Wu}
\thanks{equal contribution}
\email{yantaow@iphy.ac.cn}
\affiliation{%
Institute of Physics, Chinese Academy of Sciences, Beijing 100190, China
}
\author{Wen-Yuan Liu}
\thanks{equal contribution}
\email{wyliu@zju.edu.cn}
\affiliation{Institute for Advanced Study in Physics, Zhejiang University, Hangzhou 310027, China}

\date{\today}

\begin{abstract}

We propose a novel tensor network method to achieve accurate and efficient simulations of Abelian lattice gauge theories (LGTs) in (2+1)D \change{ for both ground state and real-time dynamics}.
The first key is to identify a gauge canonical form (GCF) of gauge-invariant tensor network states, 
which already simplifies existing algorithms for (1+1)D LGTs.
The second key is to employ the GCF of projected entangled-pair state (PEPS) combining with variational Monte Carlo (VMC),
\change{enabling efficient computations for (2+1)D LGTs}.
We demonstrate the versatile capability of this approach for accurate \change{ground state} simulation of pure $\Z_2$, $\Z_3$ and $\Z_4$  gauge theory, odd-$\Z_2$ gauge theories, and $\Z_2$ gauge theory coupled to hard-core bosons, on square lattices up to $32\times 32$. \change{Furthermore, we demonstrate that it allows for accurate simulations of real-time dynamics up to long-time, exemplified by the dynamics of elementary excitations of the deconfined $\Z_2$ gauge field on a $10\times 10$ lattice.
This is also the first example of using VMC to simulate the real-time dynamics of PEPS, whose impact may extend beyond gauge theory. 
}

\end{abstract}

\maketitle
\textit{Introduction.} The study of lattice gauge theories (LGTs) constitutes a cornerstone in modern physics.
They play foundational roles in quantum chromodynamics for studying quark confinement and hadron structure~\cite{wilson1974,kogut1979,Fukushima2011}, and also provide critical insights into condensed matter physics, where low-energy effective theories of strongly correlated systems such as quantum spin liquids and topological orders have gauge structures~\cite{wen2004quantum,kitaev2006anyons}.
The traditional Monte Carlo sampling of partition functions is a very successful computational paradigm for LGTs, however, \change{it remains challenging for equilibrium properties plagued by sign problem~\cite{troyer2005} and real-time dynamics.}
These limitations have spurred intense efforts to develop 
\change{computation in the Hamiltonian formulation,} such as quantum simulations~\cite{gross2017quantum,bruzewicz2019trapped,krantz2019quantum,yang2020observation,banuls2020simulating}.

Tensor network states (TNS) have emerged as a promising, sign-free classical simulation approach for LGT~\cite{Tagliacozzo2011entangle,Tagliacozzo2014tensor,buyens2014matrix,Silvi2014,zou2014,haegeman2015gauging,kuramashi2019three,emonts2020gauss,felser2020,magnifico2021lattice,meurice2022,cataldi2024}.
In (1+1)D, TNS in the form of  matrix product state (MPS), has been established as a reliable numerical methodology~\cite{LGT_DMRG1,LGT_DMRG2,bamuls2013the,rico2014tensor,buyens2014matrix,banus2020review}. 
Extending to (2+1)D, projected entangled pair state (PEPS)~\cite{PEPS2004} provides a compelling theoretical framework of LGTs~\cite{Tagliacozzo2014tensor,rico2014tensor,haegeman2015gauging,zohar2015fermion,zohar2016building,Blanik2024,Roose2024}. Nevertheless, PEPS-based simulations face substantial challenges stemming from both the intrinsic complexity of higher-dimensionality and the rigorous requirements of gauge constraints. 
\change{Recent explorations using gauge invariant Gaussian PEPS~\cite{zohar2018combining,emonts2023finding,emonts2020varational,kelman2024gauged} and non-gauge-constrained PEPS~\cite{iPEPS_LGT2021} have made first attempts for ground state simulations of pure $\Z_2$ and $\Z_3$ LGTs. Nonetheless, advancing versatile and high-precision PEPS methodologies capable of tackling generic LGTs remains a crucial objective for both ground state and dynamical studies, given the inherent power of PEPS to characterize strongly correlated quantum matter.}

\begin{figure}[hbt]
    \centering
     \begin{minipage}[t]{0.48\linewidth}
        \centering
        \includegraphics[scale=1]{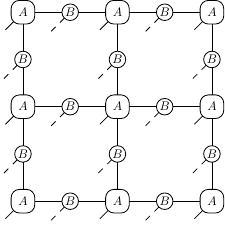} 
        \vspace{0.2cm} 
        \label{fig:lgt_peps}
    \end{minipage}%
    \hfill
    \begin{minipage}[t]{0.48\linewidth}
        \centering
        \vspace{-4cm}
        \includegraphics[scale=1]{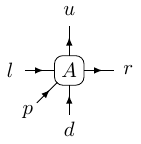} 
        \label{fig:A_tensor}
        \vspace{0.27cm} 
        \includegraphics[scale=1.5]{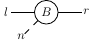} 
        \label{fig:B_tensor}
    \end{minipage}

    \caption{Left is a diagram of a gauge-invariant PEPS for LGTs. Right are the matter tensor $A$ and gauge tensor $B$. Here $d$ is the down-leg of $A$, and not the spatial dimension.}
    \label{fig:main}
\end{figure}

In this work, \change{adopting the Hamiltonian formulation}, we develop a PEPS-based computational framework to achieve accurate simulations of a wide range of (2+1)D Abelian LGTs.
A key element is the identification of a gauge canonical form (GCF) for gauge-invariant (GI) TNS, with which we can already significantly simplify MPS-based methods in (1+1)D.
In (2+1)D, the GCF enables a particularly efficient treatment of GI-PEPS via variational Monte Carlo (VMC), facilitating precise ground-state \change{and dynamical} calculations.
Our method is extensively validated with \change{ground state} simulations of $\Z_2$, $\Z_3$ and $\Z_4$ pure gauge theory, odd-$\Z_2$ gauge theory, and $\Z_2$ gauge theory coupled to hard-core bosons, on square lattices up to $32\times 32$, 
\change{and dynamical simulation of vison propagation in the deconfined $\Z_2$ gauge field.}
These results establish PEPS as a powerful pathway for accurate TNS simulations of (2+1)D Abelian LGTs, providing a new tool for non-perturbatively studying LGTs and benchmarking quantum simulations.

\textit{Hamiltonian.}  We briefly review the LGT Hamiltonians~\cite{kogut1975, emonts2020gauss}.
A ($d$+1)D LGT is defined on a $d$-dimensional cubic lattice, with gauge fields on the links and  matter fields on the vertices. 
For an Abelian gauge group $\Z_N$ (or $U(1)$), the link Hilbert space is spanned by the eigenstates $\ket{n}$ of an electric field operator $E$ such that $E\ket{n} = n \ket{n}, n \in \Z_N \, (\text{or } \Z)$.   
Its raising operator $U\equiv e^{i\phi}$ is the exponential of its canonical conjugate operator $\phi$: $[\phi, E] = i$ and $U \ket{n} = \ket{n+1 \text{ mod } N}$. 
The matter Hilbert space hosts a boson or a fermion on the vertex $\vec x$ with annihilation operator $c_{\vec x}$.  
The gauge invariance, at every $\vec x$, is enforced as  
\begin{equation}
  c^\dag_{\vec x}c_{\vec x} + \sum_{\alpha=1}^d (E_{(\vec x-\vec e_\alpha,\alpha)} - E_{(\vec x, \alpha)}) = Q_{\vec x}  \text{ mod } N,
\label{eq:gauss_N}
\end{equation}
where $(\vec x, \alpha)$ is a link between $\vec x$ and $\vec x + \vec e_\alpha$, with $\vec e_\alpha$ being the unit vector along the $\alpha$-th axis.
$Q_{\vec x}$ is a preset integer, fixing the gauge sector.
The LGT Hamiltonian is: 
\begin{align}
  &H = H_M + H_B + H_E ,
\\
 & H_M = \sum_{\vec x} m_{\vec x}c^\dag_{\vec x} c_{\vec x} + \sum_{\vec x, \alpha} (Jc_{\vec x}^\dag U_{(\vec x, \alpha)} c_{\vec x + \vec e_\alpha} + h.c.) ,
  \label{eq:matter}
\\
&H_B = -h \sum_{\vec x} U_{\vec x, 1} U_{\vec x+\vec e_1,2} U^\dag_{\vec x+\vec e_2,1}U^\dag_{\vec x,2} + h.c. ,
\label{eq:H_b}
\\
 & H_E = g \sum_{\vec x, \alpha} 2-2\cos(2\pi E_{(\vec x, \alpha)}/N) \text{ or } g\sum_{\vec x, \alpha} E_{(\vec x, \alpha)}^2,
  \label{eq:electric}
\end{align}
where $m_{\vec x}$ is the chemical potential (or the bare particle mass), and $J$ is the gauge-matter coupling strength.
$H_B$ (present only for $d\ge2$) and $H_E$ are respectively the magnetic and electric energy terms. 
The two instances of $H_E$ are for either the $\Z_N$ or $U(1)$ gauge group. 

\textit{Gauge-invariant TNS.} 
Gauge-invariant tensor network states naturally describe the physical Hilbert space of LGTs~\cite{Tagliacozzo2011entangle,Tagliacozzo2014tensor,buyens2014matrix, rico2014tensor,Silvi2014,haegeman2015gauging,zohar2015fermion,zohar2016building}.
To construct a GI-TNS wavefunction, one works in the basis of particle occupation number and electric fields.
As in Fig.~\ref{fig:main}, the network has three‐leg $B$ tensors for gauge fields and  ($2d+1$)-leg $A$ tensors for matter fields. 
Gauge invariance of the wavefunction is enforced by imposing sparsity constraints on $A$ and $B$.
Specifically, we assign charges $q(j)$ in $\Z_N$ (or $\Z$) to tensor indices $j$ on each virtual leg, then in two spatial dimensions, tensor blocks of $A$ and $B$ satisfy \cite{buyens2014matrix}
\begin{align}
    & A^p_{lrdu} = {\mathcal{A}}^p_{lrdu} \delta_{p + q(l) + q(d) - q(r) - q(u), \, Q_{\vec x}},
    \label{eq:A} \\
    & B^n_{lr} = \mathcal{B}^n_{lr} \delta_{n, q(l)}\delta_{n, q(r)} .
    \label{eq:B}
\end{align}
The bond dimension of the TNS is then $D = \sum_k D_k$, where $D_k$ is the degeneracy of the charge sector $k$, i.e. the number of tensor indices $j$ with $q(j) = k$.
Although this GI ansatz has been known for a decade, and its MPS algorithms has been established in (1+1)D~\cite{buyens2014matrix}, the algorithmic feasibility of GI-PEPS including optimization for ground states and \change{real-time dynamics} and computation of physical quantities, has remained a major roadblock, preventing the power of PEPS from manifesting for (2+1)D LGTs.
Below we show how to overcome these challenges.

\textit{Gauge canonical form.} 
A key ingredient for our approach is the GCF, which we now identify.  
On the link connecting an $A$ tensor and a $B$ tensor, one can define the following block-diagonal matrix: 
\begin{equation}
  X = \bigoplus_{k \in \Z_N (\text{or } \Z)} \mathcal{B}^{[k]}
\end{equation}
where $\mathcal{B}^{[k]}$ is a $D_k\times D_k$ matrix obtained from choosing $n=k$ in $B_{lr}^n$ and restricting to the $l$ and $r$ indices with charges equal to $k$. 
Using gauge transformations $A \rightarrow AX, B \rightarrow X^{-1}B$, the gauge tensor $B$ simplifies as
\begin{equation}
  B_{lr}^n = \delta_{lr} \delta_{n, q(l)} \delta_{n,q(r)},  
  \label{eq:gcf}
\end{equation}
and $A$ keeps the same form as Eq.(\ref{eq:A}). 
We refer to this new form as the GCF, in which the $B$ tensors no longer contain variational parameters and one only needs to optimize the $A$ tensors.
Below we show how GCF enables efficient computations of GI-MPS (see End Matter) and GI-PEPS. 

\textit{VMC.}
In 2D, GI-PEPS simulations are challenging due to their intrinsic complexity, gauge-invariance constraints, and the four-body plaquette terms in the Hamiltonian [Eq.~(\ref{eq:H_b})]~\cite{iPEPS_LGT2021}.
We find that combining VMC and GCF overcomes these challenges effectively.
In VMC, the expectation value of an observable is calculated as $\braket{O} = \sum_{\vec s} \frac{\abs{\braket{\vec s|\Psi}}^2}{\braket{\Psi|\Psi}} \frac{\braket{\vec s|O|\Psi}}{\braket{\vec s|\Psi}}$, where $\ket{\vec s}$ labels a configuration of gauge and matter fields.
This sum is estimated via sampling $\ket{\vec s}$ from the probability distribution $\frac{\abs{\braket{\vec s|\Psi}}^2}{\braket{\Psi|\Psi}}$, where the basic component is evaluating single-layer networks $\braket{\vec s|\Psi}$~\cite{sandvik2007,schuch2008,wang2011,liu2017,liu2021}.
We only sample physical configurations, as unphysical ones' amplitudes are exactly zero in GI-PEPS. 

The sampling and GCF critically simplifies computations: each configuration $\ket{\vec s}$ uniquely selects a single charge sector of matter tensors $A$ with gauge tensors $B$ absent.
Thus, tensors in the resulting network $\braket{\vec s|\Psi}$ only have bond dimension $D_k$, significantly reduced from the total bond dimension $D=\sum_{k=1}^{N}D_k$ for $\Z_N$ gauge group. 
This allows efficient computations using advanced PEPS-VMC techniques~\cite{liu2017,liu2021,liu2024tnf,liu2025} that have been used to study frustrated spin systems~\cite{liu2018gapless,liu2022gapless,liu2022emergence,liu2024emergent,liu2024j1j2j3,liu2024quantum}.
See SM~\cite{SM} for more discussions.

We use gradient-based stochastic reconfiguration~\cite{SR1998,tvmc,directSam2021,liu2025} \change{for both ground state and real-time dynamics \cite{tvmc1,tvmc2,tvmc3} }of GI-PEPS for systems on the open square lattice.
The computational cost scales as $O(D_k^5\chi^2+D_k^4\chi^2+D_k^3\chi^3)$, dominated by plaquette term evaluations and variational boundary MPS compression.
$\chi$ is the cutoff bond dimension of the boundary MPS for contracting $\braket{\vec s|\Psi}$, with $\chi=3D_k$ being good enough.
Then the energy measurement scales as $O(MN_{\rm site} D_k^7)$, where $N_{\rm site}$ is the  size and  $M$ is the number of Monte Carlo sweeps~\cite{SM} typically on the order of $10^4$ with statistical uncertainty about $10^{-5}$.

\textit{Pure $\Z_N$ gauge theory.} 
We first consider pure $\Z_2-\Z_4$ gauge theories (no matter field, $Q_{\vec x} \equiv 0$).
We use $D_k=2$ which we find is good enough for convergence on relevant sizes (see SM~\cite{SM}).
Fixing $h=1$, we scan $g$ to compute ground state properties. For the well understood $\Z_2$ pure gauge theory, the ground state energies computed with GI-PEPS on the $16\times 16$ lattice, agree excellently with those of Quantum Monte Carlo (QMC), showing high accuracy up to $10^{-5}$. See more results in SM \cite{SM}.

\begin{figure}[tbp]
 \centering \includegraphics[width=0.48\textwidth]{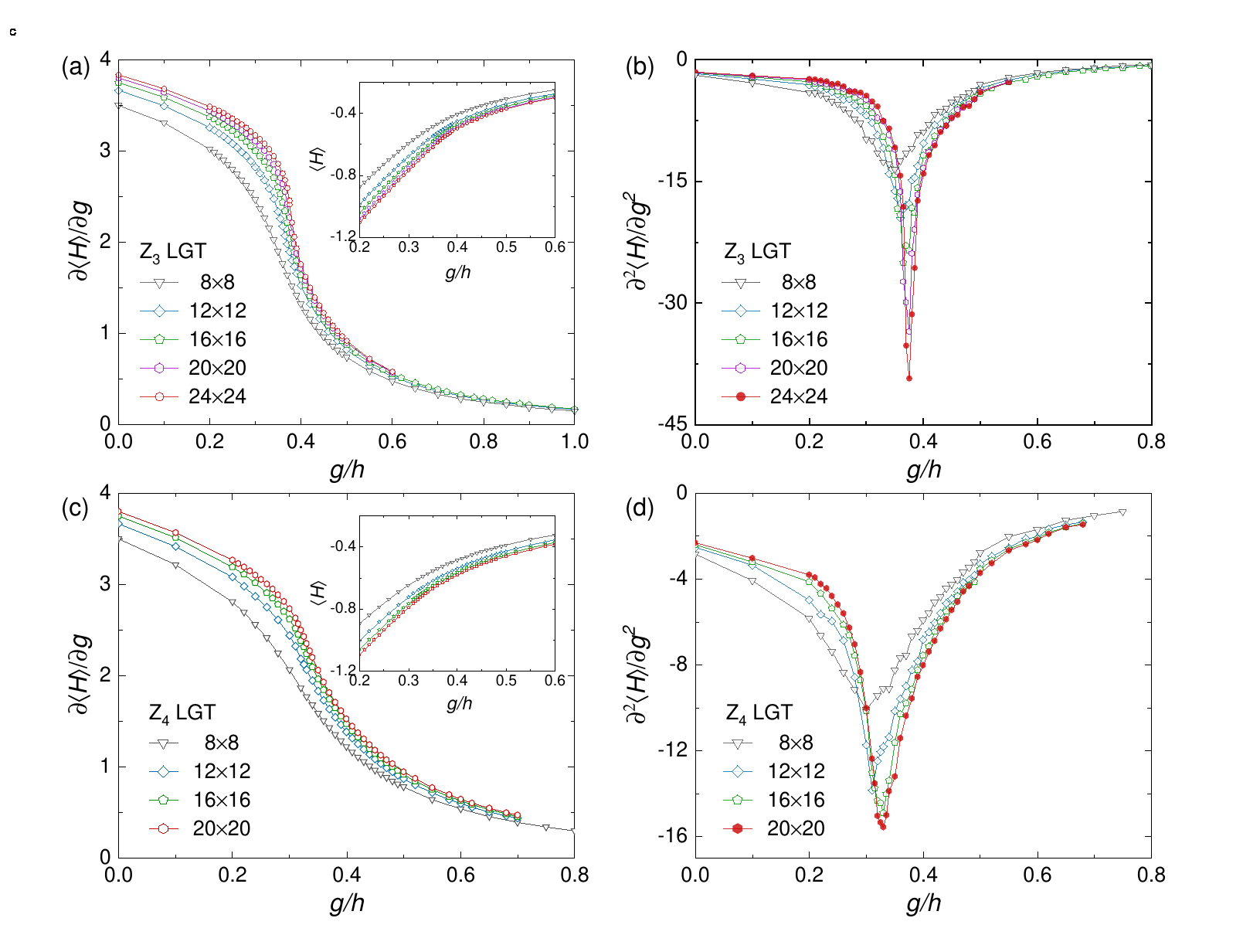}
 \caption{Results of $\Z_3$ (a-b), $\Z_4$ (c-d) LGTs at various $g$, including ground state energy $\langle H \rangle $ [insets of (a) and (c)], the first-order and second-order energy derivative.  
 }
 \label{fig:pureZn}
 \end{figure}

\begin{figure*}[tbp]
 \centering
  \includegraphics[width=\textwidth]{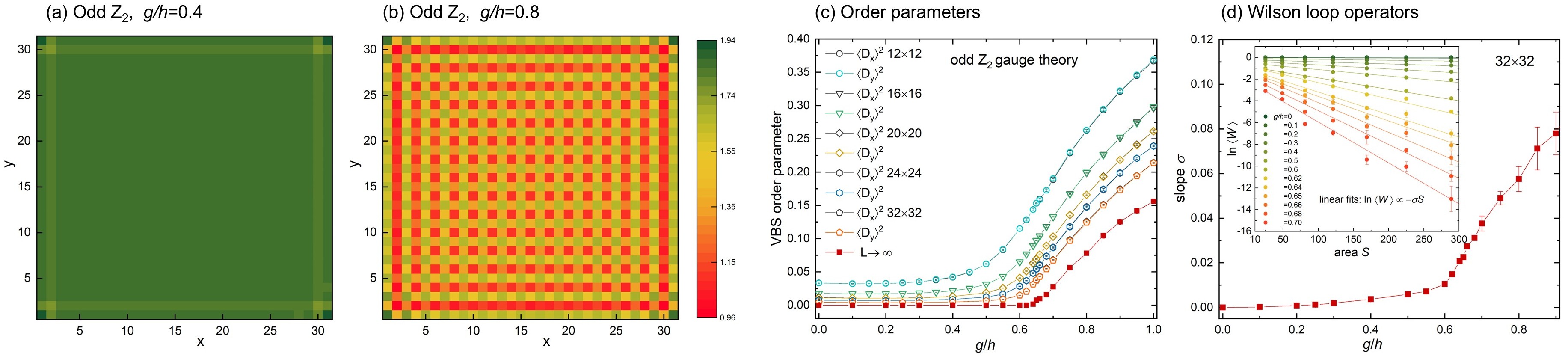}
 \caption{Results of odd $\Z_2$ LGT. (a) and (b) present the plaquette value  $\langle P_{\vec x} \rangle$ at each site $\vec{x}$ on $32\times 32$  at $g/h=0.4$ and 0.8.  (c) shows the  VBS order parameters $\langle D_x\rangle ^2 $ (black) and $\langle D_y\rangle ^2 $ (colorful), and  red symbols are the values in thermodynamic limit extrapolated using quadratic fits of $\langle D_x\rangle ^2 $. The inset of (d) is the linear-linear plot of ${\rm ln} \braket{W}$ versus area $S$ (different central regions on $32\times 32$) to extract  $\sigma$ following $\braket{W} \propto {\rm e}^{-\sigma S}$; the main panel shows the $g-$dependence of  the slope $\sigma$.  }
 \label{fig:odd_Z2}
 \end{figure*}

For the $\Z_3$ case, we compute ground-state properties for sizes from $8\times 8$ to $24\times 24$.
The first derivative of energy, $\frac{\partial \langle H\rangle}{\partial g}=\frac{1}{g}\langle H_E\rangle$ [Fig.~\ref{fig:pureZn}(a)], and its finite-difference second derivative $\frac{\partial^2 \langle H\rangle}{\partial g^2}$ [Fig.~\ref{fig:pureZn}(b)], reveal clear signatures of a first-order transition, consistent with early studies~\cite{Bhanot1980}. The transition point from small sizes shows a minor shift. The convergence between $20\times 20$ and $24\times 24$ yields a thermodynamic-limit critical point at $g_c=0.375(3)$.
This agrees well with a recently found $g_c = 0.37(1)$ using neural quantum states on torus up to size $10\times 10$~\cite{Apte2024deep}.
This is to be compared with the non-gauge-constrained iPEPS result $g_c=0.448(3)$~\cite{iPEPS_LGT2021},
\change{whose ground state energies obtained near the critical region, are higher than our results extrapolated to infinite size (see SM~\cite{SM}).}
The quantitative difference may arise from the reliance of iPEPS on simple update optimization rather than the fully variational optimization employed here, demonstrating the challenge of such calculations.

For the $\Z_4$ case, unexplored previously by TNS, we extend our analysis to $20\times 20$ sites.
As shown in Figs.~\ref{fig:pureZn}(c,d), the energy derivatives suggest a phase transition at $g_c = 0.330(5)$,
by comparing results from size $16\times 16$ and $20\times 20$. This constitutes the first PEPS study of $\Z_4$ LGT, offering a benchmark for higher-order gauge groups. Very interestingly, this $g_c$ overlaps with that of the $\Z_2$ theory, 0.3285.
In fact, for the 3D classical gauge theories, the $\Z_4$ theory is proven to be equivalent to the $\Z_2 \times \Z_2$ theory~\cite{grosse1981equivalence}.
While no proof is available in the quantum case, we {\it conjecture} the equivalence here. 

\textit{Odd-$\Z_2$ theory.}
Another representative example is the odd $\Z_2$ gauge theory, i.e. with $Q_{\vec x} \equiv 1$ for all $\vec x$, relevant for understanding spin liquids and dimer models~\cite{sachdev1991,senthil2000,fradkin2001,Sachdev2019}. 
According to theoretical predictions~\cite{senthil2000,fradkin2001,Sachdev2019}, by varying $g$ it experiences a continuous transition between a deconfined phase and a confined phase that breaks translation symmetry.
Its dual model - the fully frustrated transverse field Ising model~\cite{fradkin2001}, has been studied by QMC~\cite{FFTFIM2012}.
With GI-PEPS, we are able to directly obtain its ground state \textit{wavefunction}.
Figs.~\ref{fig:odd_Z2}(a) and (b) show the plaquette operator $P_{\vec x}=U_{\vec x, 1} U_{\vec x+\vec e_1,2} U^\dag_{\vec x+\vec e_2,1}U^\dag_{\vec x,2} + h.c.$ on a $32\times 32$ lattice, revealing a uniform and a symmetry broken phase at $g=0.4$ and $0.8$, respectively. 

To precisely locate the transition point, we compute the valence-bond solid (VBS) order parameter~\cite{liu2022emergence}
\begin{equation}
D_{x/y} = \frac{1}{L(L-1)}\sum_{\vec x} (-1)^{x_\alpha}\bar{E}^{\alpha}_{\vec x},  
\end{equation}
where $\alpha=1, 2 ~{\rm for}~ D_{x}, D_{y}$. 
$\bar{E}^{\alpha}_{\vec x}=2-2\cos(\pi E_{(\vec x, \alpha)})$ is the electric field strength on the link $(\vec x, \alpha)$ [see Eq.(\ref{eq:electric})],  and $\vec{x}=(x_1,x_2)$ is the vertex position.
Note the identical $\langle D_x\rangle ^2 $ and $\langle D_y\rangle ^2 $ in Fig. \ref{fig:odd_Z2}(c), which reflects the $C_4$ rotation symmetry.
Through quadratic finite-size extrapolation, we obtain VBS order parameters in the thermodynamic limit [red curve in Fig. \ref{fig:odd_Z2}(c)], locating the phase transition point at $g_c=0.64(1)$, in good agreement with the QMC results $g_c\simeq0.634$ from the dual model~\cite{FFTFIM2012}.

The  Wilson loop operator $\braket{W}$ on the $32\times 32$ lattice is shown in Fig. \ref{fig:odd_Z2}(d).
The slope $\sigma$, extracted from  $\braket{W} \propto {\rm e}^{-\sigma S}$, remains small for $g\lesssim 0.6$ but increases sharply afterward, signaling a perimeter-law to area-law transition consistent with $g_c = 0.64(1)$. These nonlocal operator calculations complements those of local operators $D_{x/y}$ in detecting translation symmetry breaking, providing insight into the underlying physics from the perspective of deconfinement.

\begin{figure}[tbp]
 \centering
  \includegraphics[width=0.47\textwidth]{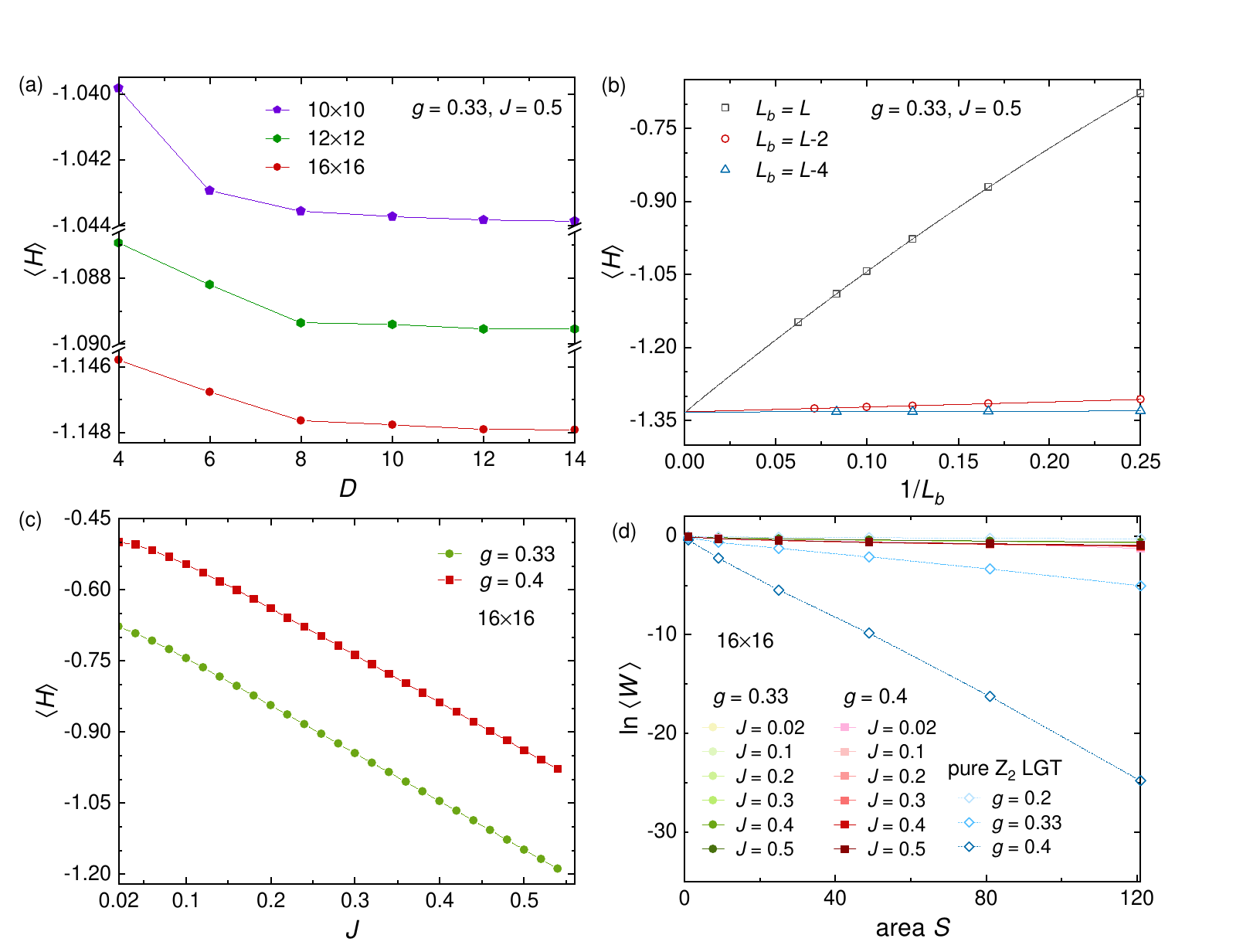}
 \caption{Results of $\Z_2$ gauge field coupled to hard-core bosons at half-filling ($h=1$). (a) Convergence of energy with respect to bond dimensions. (b) Finite-size scaling of the energy using central bulk $L_b\times L_b$ energy of an $L\times L$ lattice. Quadratic fits are used for extrapolations. (c) $J-$dependence of energy at $g=0.33$ and $0.4$ on $16\times 16$ lattice. (d) Wilson loop operators on $16\times 16$ lattice at different $J$ for a given $g=0.33$ (green) and $g=0.4$ (red), compared to the pure $\Z_2$ LGT (blue). The green and red lines are respectively largely overlapped, both very close to the pure $\Z_2$ LGT at $g=0.2$ (lightest blue).}
 \label{fig:Z2_boson}
 \end{figure}

\textit{$\Z_2$ gauge theory coupled to hard-core bosons.} 
We demonstrate that we can directly deal with matter fields.  
Here we consider $\Z_2$ gauge fields coupled to hard-core bosons. 
Its (1+1)D version has been studied, known as the $\Z_2$ Bose-Hubbard model \cite{Z2BH_2019}, while the (2+1)D case remains uncharted. 
For benchmarking with exact diagonalization (ED) calculation, we first consider a $3\times 3$ square lattice with 2 bosons.
The definite boson number is realized by sampling in the corresponding particle number subspace.
Taking $(h,g,J)=(1,0.33,0.5)$, the optimized $D=6$ PEPS gives the energy persite $-0.470713502(3)$ using $M=10^5$ samples, matching the ED energy $-0.4707135061$ excellently.  

We then scale up to $16\times 16$ sites at half filling of bosons.
Fig.~\ref{fig:Z2_boson}(a) presents the energies from PEPS with bond dimensions $D$ up to $14$ for different sizes at $(h,g,J)=(1,0.33,0.5)$.
Unlike the pure $\Z_2$ LGT where $D=4$ is sufficient, the matter-coupled case requires $D=12$.
These results reflect the increased entanglement of this model and our ability to handle large bond dimensions.
We also compare the thermodynamic-limit energy evaluated using different central bulk energies for extrapolations~\cite{liu2021,liu2022emergence}.
Shown in Fig.~\ref{fig:Z2_boson}(b), given a central bulk region of $L_b\times L_b$~\cite{SM}, for example, $L_b=L-2$, the extrapolated energy for the thermodynamic limit is $-1.3322(4)$, in good agreement with those from other choices of $L_b=L$ and $L_b=L-4$ that are $-1.3337(4)$ and $-1.3322(2)$, respectively.
This consistency corroborates our results~\cite{liu2021,liu2022emergence}.

 One also expects that, in the presence of dynamical matter fields, the Wilson loop operator exhibits a perimeter-law even in the confinement regime of the pure $\Z_2$ LGT, due to screening by the matter field~\cite{Fradkin2013}. 
This is indeed what we observe. We present the energy and Wilson loop operator of $16\times 16$ lattice in Figs.~\ref{fig:Z2_boson}(c) and (d). For pure $\Z_2$ LGT, as shown previously, $g=0.2$, $0.33$ and $0.4$ correspond to the deconfined, near critical and confined regimes, respectively. From Fig.~\ref{fig:Z2_boson}(d) we see after adding matter fields, at $g=0.33$ and $0.4$, Wilson loop operators for different $J$ show perimeter-law behavior.

\change{
\textit{Real-time dynamics in (2+1)D LGTs.} A key advantage of the Hamiltonian formulation lies in its natural suitability for studying real-time dynamics. Here, we directly demonstrate this capability by simulating the real-time evolution of a $\mathbb{Z}_2$ LGT using GI-PEPS.
We use the scheme of time-dependent VMC (tVMC) \cite{tvmc1,tvmc2}, which recently have been applied to neural quantum states \cite{nqstvmc_1,nqstvmc_3,nqstvmc_4,nqstvmc_5,nqstvmc_6,nqstvmc_7,nqstvmc_8} and we adapt these techniques within our GI-PEPS framework. Notably, this work presents the first application of tVMC to PEPS, an advancement with potential implications far beyond gauge theory.

As a concrete example, we investigate the dynamics of a vison—an excitation of the deconfined $\mathbb{Z}_2$ gauge field corresponding to a violation of the plaquette operator $P_{\rm x}$ \cite{senthil2000}. 
We initialize a vison at the bottom-left of the lattice over the deconfined ground state ($g = 0.1$) and evolve it under the same parameter, with a time step $\Delta t=0.005$.
On a $6 \times 6$ lattice, GI-PEPS reproduces the exact diagonalization results with high accuracy, as seen in the expectation values $\braket{P_{\rm x}}$ at selected plaquettes (Fig. \ref{fig:vison}a). On a $10 \times 10$ lattice, a system far beyond the reach of exact methods (Hilbert space dimension $2^{81}$), the vison propagates cleanly toward the top-right [Figs. \ref{fig:vison}(b-d)], in a simulation time up to $T = 18$ (7200 evolution steps using second order expansions~\cite{SM}), while maintaining energy conservation within 0.2\% \cite{SM}. This showcases the power of our approach in capturing long-time, large-scale quantum dynamics. See the SM \cite{SM} for a movie of the vison propagation and the details of the tVMC.

\begin{figure}[hbt]
    \centering
    \begin{minipage}[t]{0.48\linewidth}
        \centering
        \includegraphics[width=\linewidth]{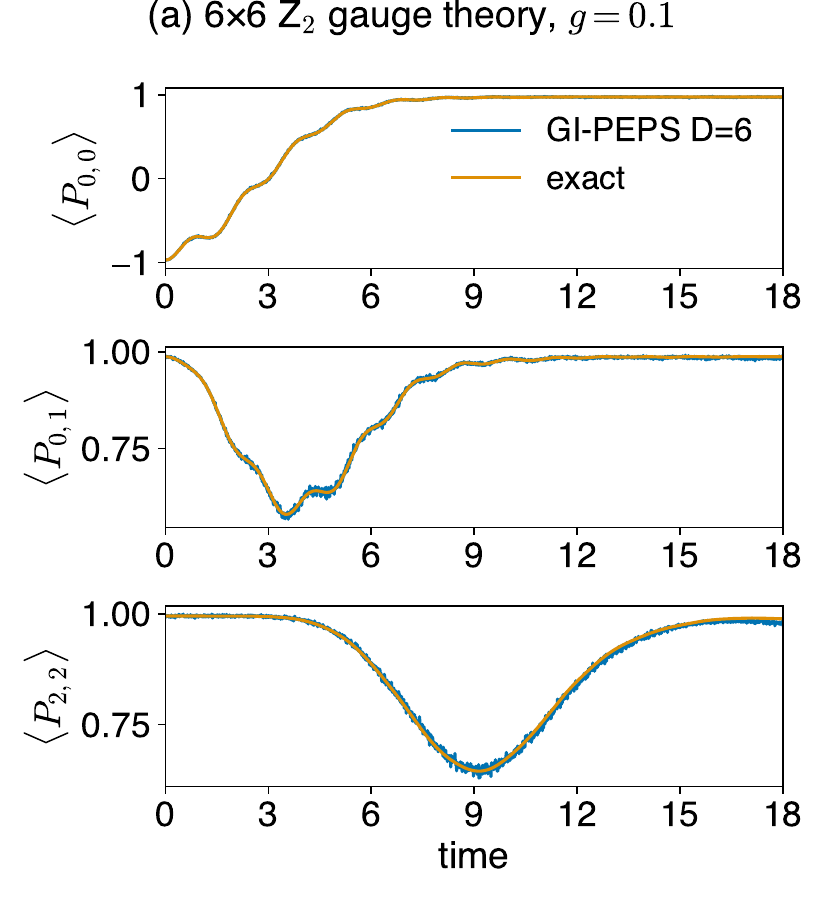}
    \end{minipage}%
    \hfill
    \begin{minipage}[t]{0.48\linewidth}
        \centering
        \includegraphics[width=\linewidth]{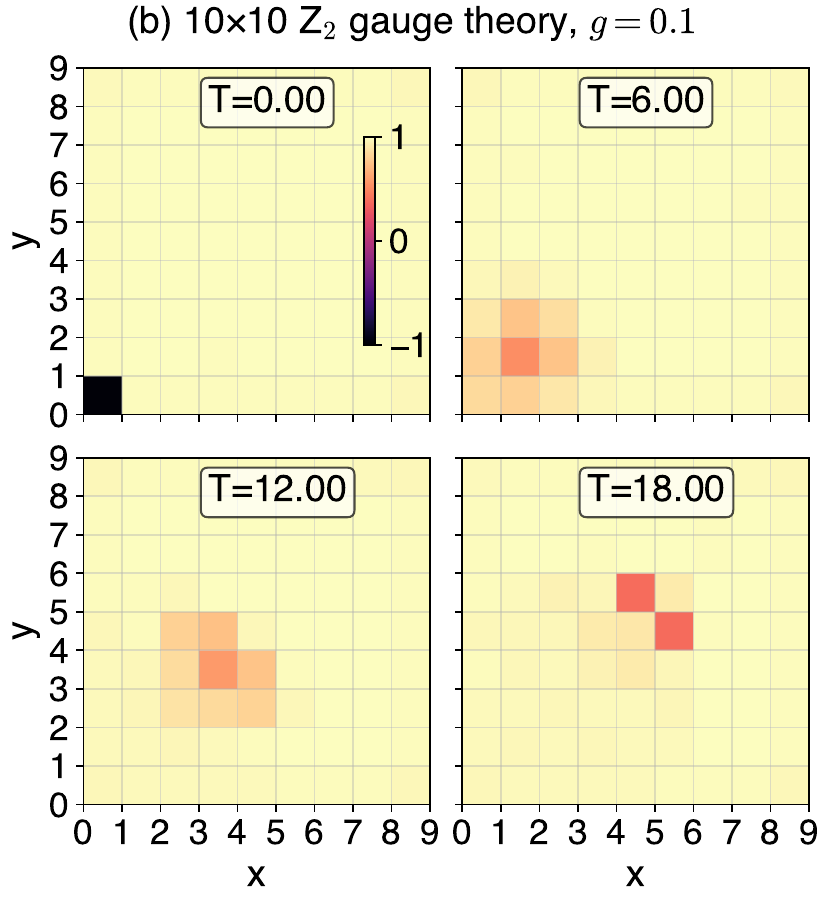}
    \end{minipage}
    \caption{ Vison dynamics. The initial state is prepared by acting the Pauli matrix $\sigma_z$ in the electric basis $\{\ket{0},\ket{1}\}$ on the gauge field living on the bottom-left vertical link over the corresponding ground state, in order to create a vison excitation.
    (a) Real-time dynamics of a vison in a $6\times6$ lattice ($D=6$), compared with exact diagonalization. 
    (b) Larger-scale simulation on $10\times10$ lattice showing vison propagation ($D=6$ and $D=8$ give the same result \cite{SM}).
    }
    \label{fig:vison}
\end{figure}
}

\textit{Conclusion and discussion.} 
We have developed a powerful gauge-invariant tensor network computational framework for (2+1)D Abelian LGTs, overcoming longstanding challenges in algorithmic feasibility of GI-PEPS. 
 Central to our approach is the gauge canonical form, which fixes gauge tensors to parameter-free forms- thereby reducing variational parameters exclusively to matter tensors, as well as its further combination  with VMC.
 We validate the framework across diverse models, achieving accurate large-scale simulations up to $32\times 32$ sites for ground states and $10 \times 10$ for dynamics. 
\change{ These advances establish gauge-invariant PEPS as a state-of-the-art tool for both ground-state and real-time dynamics simulations of Abelian LGTs. This framework opens up a new avenue for non-perturbative studies, and the GCF will be also potentially inspiring for non-Abelian cases. We believe this work will have a foundational impact on Hamiltonian-based approaches to LGTs. Generalization to fermionic matter is reserved for future work.
}

The potential of GI-PEPS and GCF remains to be seen for double-layer TNS methods \footnote{We would like to thank Jutho Haegeman for sending us a recent unpublished master thesis from his group, where exact gradient descent using automatic differentiation was attempted for finding the ground states of an infinite-PEPS with structure similar to GCF. 
Although the numerical results are yet to converge with the bond dimension in the thesis, they do show the possibility  of such calculations using more conventional methods based on double-layer TNS contraction.
}. Another possibility recently proposed is to modify the GI-PEPS ansatz as a symmetric PEPS by embedding the gauge group $G$ into an enlarged globally symmetric theory with group $G\times G$~\cite{canals2024}.

\begin{acknowledgements}
\textit{Acknowledgement.} 
Y.W. would like to thank discussion with Akira Matsumoto, Etsuko Itou, Masazumi Honda, and Tetsuo Hatsuda during the early stage of the work. W.Y.L. is especially grateful to Garnet K. Chan for encouragement. 
Y.W. was supported by the RIKEN iTHEMS fellowship, and is currently supported by  a start-up grant from IOP-CAS. 
W.Y.L. is supported by a start-up grant from Zhejiang University.
Part of the code is tested using the TenPy code base \cite{hauschild2018efficient, johannes2024tensor}.
\end{acknowledgements}

\section{End matter}
\textit{GCF in (1+1)D.} 
In the End Matter, we demonstrate that the GCF enables an elegant algorithm of the time evolution block decimation (TEBD) \cite{TEBD} for LGT in (1+1)D. 
We take the Schwinger model as an example, which is a toy-model of quantum electrodynamic in (1+1)D \cite{schwinger1962}.
Its Hamiltonian is $H_M + H_E$ with $U(1)$ gauge group and staggered fermions $Q_x = (1+(-1)^x)/2$ and $m_x = (-1)^xm$.  

Due to the GCF, simulations do not explicitly require the gauge tensors. 
Instead, the time evolution of the Schwinger model is simulated via a $U(1)$-symmetric TEBD applied to the Hamiltonian with global symmetry:
\begin{equation}
H = \sum_x m_x c_x^\dag c_x + \sum_x J c_x^\dag c_{x+1} + h.c., 
\label{eq:H_globalsymm}
\end{equation}
with an MPS made only of $A$ tensors.
The would-be gauge canonical $B$ tensors dictate that the virtual charges of $A$ encode the physical electric fields.
Instead of using Jordan-Wigner transformation, we directly use the fermionic MPS in the swap gate formalism \cite{shi2009graded,PhysRevB.81.165104}.

Simulating Eq. (\ref{eq:H_globalsymm}) is different from the LGT systems in three aspects: 
\begin{enumerate}
\item
The electric part $H_E=g\sum_{x} E_x^2$ is missing.
\item 
The hopping $c_x^\dag c_{x+1}$ is different from the gauge-invariant hopping $c_x^\dag U_x c_{x+1}$. 
\item 
For systems with global symmetry, there is an arbitrariness in the symmetry charge of the MPS tensors $A_x$. 
Suppose one has tensor $A_x$ and $A_{x+1}$, with symmetry charge $Q_x$ and $Q_{x+1}$. 
Combining $A_x$ and $A_{x+1}$ gives a two-site tensor $\Lambda \equiv A_{x}A_{x+1}$, with symmetry charge $Q_x+Q_{x+1}$.  
The TEBD gate tensor $U$ changes $\Lambda$ to $\Lambda'$, but does not change the symmetry charge of $\Lambda$. 
When one splits $\Lambda'$ to $A'_x$ and $A'_{x+1}$ via singular value decomposition, one can assign any symmetry charge $Q'_x$ to $A'_x$ as long as $A'_{x+1}$'s symmetry charge is then assigned as $Q_{x} + Q_{x+1} - Q'_x$, and the virtual charge on the link between $x$ and $x+1$ properly redefined.
This arbitrariness is absent for LGTs, as the symmetry charge $Q_x$ is pre-defined.
\end{enumerate}
To accommodate these differences, two additions are needed during the TEBD of Eq. (\ref{eq:H_globalsymm}): 
\begin{enumerate}
    \item  At each link, apply the time evolution gate $e^{-id\tau g E_{x}^2}$ [Eq. (\ref{eq:electric})] ($d\tau$ being the time step) on the virtual charges of $A_x$
    \item  Keep the symmetry charge of $A_x$ as $Q_x$ when splitting a two-site wavefunction $A_xA_{x+1}$ .
During the TEBD of Eq.(\ref{eq:H_globalsymm}), the hopping term $c_x^\dag c_{x+1}$ shifts the symmetry charge of $A_x$ ($A_{x+1})$ by +1 $(-1)$. 
One can exploit the gauge freedom of the symmetry charges in an MPS with global symmetry to redefine the post-split tensors' symmetry charges --- reverting $A_x$'s to $Q_x$ and $A_{x+1}$'s to $Q_{x+1}$ --- while incrementing the virtual charge between $x$ and $x+1$ by $+1$. 
This precisely implements the gauge invariant $c_x^\dag U_x c_{x+1}$, as summarized in the Fact \ref{fact1}. 
\end{enumerate}
\begin{fact}
  Let $A'_{x}$ and $A'_{x+1}$ be the result of $c_x^\dag c_{x+1}$ acting on the two-site wavefunction $A_{x}A_{x+1}$:
\begin{equation}
\input{tikz/cdagc_AA.tex}
\end{equation}
Then 
\begin{equation}
\input{tikz/cdagc_AA_noLGT.tex}
\end{equation}
where $B$ is in GCF.
\label{fact1}
\end{fact}

Another important piece of the TEBD algorithm is the MPS isometric canonical form (not to be confused with the gauge canonical form): the truncation of the two-site wavefunction must be performed at the canonical center. 
The isometric canonical form is also preserved by the CGF due to the following equation:
\begin{equation}
\input{tikz/rho_MPS.tex}
\end{equation}
provided that $B$ is in GCF. 

The approach described here significantly simplifies the algorithm of Ref. \cite{buyens2014matrix} \change{without any approximation}, which required manual gauge field truncation and blocking of $A$ and $B$ tensors.
For example, if one cuts the gauge field at $\abs{n}_\text{max}=3$, then a physical leg dimension of $14$ is needed on each site in \cite{buyens2014matrix}, while for us the physical dimension is always 2. 
In addition, our cutoff of the gauge field is based on entanglement via SVD, which seems much more natural for an MPS. 

\begin{figure}[htb]
  \includegraphics[scale=0.4]{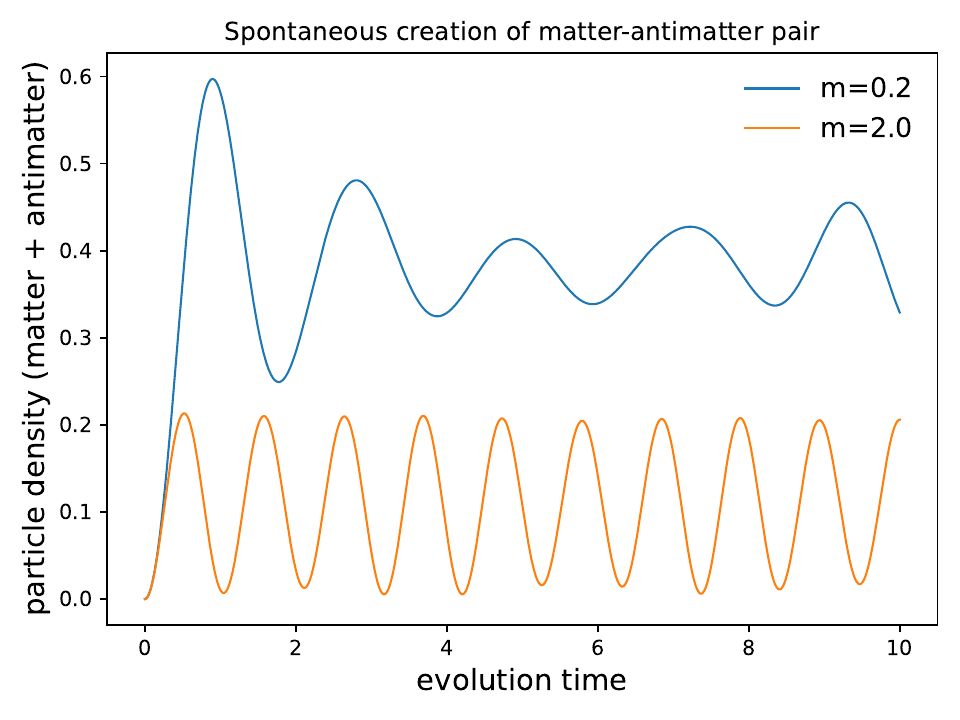}
  \caption{Schwinger mechanism on a 40-site chain. $J=i$, $g=1$.
  The time evolution is obtained using TEBD with a second-order Trotter decomposition with time step $dt = 0.01$.}
  \label{fig:schwinger}
\end{figure}

To validate our method, we perform imaginary-time evolution ($d\tau=0.01$) on a 16-site chain with $m=0.2, J=-5i,g=0.05$, and obtained ground state energy $-36.33990$, which is in excellent agreement with exact diagonalization ($-36.33994$) \footnote{Provided by Akira Matsumoto}.
Fig. \ref{fig:schwinger} further illustrates real-time evolution starting from the vacuum state: particle-antiparticle pairs are spontaneously created, and smaller fermion masses $m$ enhance particle-antiparticle pair production, directly manifesting the Schwinger mechanism \cite{schwinger1962}.

\bibliography{ref.bib}

\onecolumngrid
\appendix
\setcounter{equation}{0}
\newpage

\renewcommand{\thesection}{S-\arabic{section}} \renewcommand{\theequation}{S%
\arabic{equation}} \setcounter{equation}{0} \renewcommand{\thefigure}{S%
\arabic{figure}} \setcounter{figure}{0}

\centerline{\textbf{Supplemental Material}}

\maketitle

\section{S-1. Variational Monte Carlo }

There are standard works about the variational Monte Carlo algorithm for tensor network states~\cite{sandvik2007,schuch2008,wang2011,emonts2020varational}. It has been used in generic bosonic PEPS and fermionic PEPS for solving frustrated spin and fermionic models~\cite{liu2017,liu2018gapless,liu2021,liu2022gapless,liu2022emergence,liu2024emergent,liu2024j1j2j3,liu2024quantum,liu2024tnf,liu2025,dong2019,dong2020stable}, as well as gauged Guassian PEPS for lattice gauge theories~\cite{zohar2018combining,emonts2023finding,emonts2020varational,kelman2024gauged}.  Here we follow the presentation for generic PEPS in Ref.~\cite{liu2017,liu2021}, and discuss its application for gauge-invariant PEPS. 

 In VMC, using $\ket{\vec s}\equiv \ket{\vec n, \vec p}$ to denote gauge field configuration $\ket{\vec n}$ and matter field configuration $\ket{\vec p}$, the expectation values are computed by importance sampling of configurations $\ket{\vec s}$. For example, the total energy reads: 
\begin{equation}
  E_{\rm tot}= \frac{\langle \Psi |H|\Psi\rangle}{\langle \Psi | \Psi\rangle}=\sum_{{\bf s}}\frac{ | \langle{\bf s} \ket{\Psi}|^2}{\langle \Psi | \Psi\rangle}  \frac{\bra{{\bf s}}H|\Psi\rangle}{\bra{{\bf s}}\Psi\rangle}  
 = \sum_{{\bf s}}p({\bf s}) E_{\rm loc}({\bf s}) ~~,
 \label{eq:energy}
\end{equation}
where  $\Psi({\bf s})\equiv\bra{{\bf s}}\Psi\rangle$ is the amplitude of the configuration $|{\bf s}\rangle$, and $p({\bf s})= | \langle{\bf s} \ket{\Psi}|^2 / \langle \Psi | \Psi\rangle $ is the probability. $E_{\rm loc}({\bf s})$ is the local energy defined as  
\begin{equation}
E_{\rm loc}({\bf s})= \frac{\bra{{\bf s}}H|\Psi\rangle}{\bra{{\bf s}}\Psi\rangle}=\sum_{{\bf s}^{\prime}} \frac{\bra{{\bf s}^\prime}\Psi\rangle}{\bra{{\bf s}}\Psi\rangle} \bra{{\bf s}} H \ket{{\bf s}^\prime}.
\label{eq:Es}
\end{equation}

The sampling process in Eq.(\ref{eq:energy}) employs the conventional Markov Chain Monte Carlo (MCMC) method. Beginning with the initial configuration  $\ket{{\bf s}_{0}}$, a trial configuration  $\ket{{\bf s}_{1}}$ is proposed. According to the Metropolis algorithm, this candidate configuration  $\ket{{\bf s}_{1}}$ is accepted as the new configuration for the Markov Chain if a uniformly random number in [0,1)  is below the probability ratio $p({\bf s}_{1})/p({\bf s}_{0})$.  If rejected, another configuration state $\ket{{\bf s}_{1}^{\prime}}$ is proposed instead.

Monte Carlo sampling provides a direct framework for calculating energy gradients to be used for updating variational parameter $\alpha_m$ of the wave function:
\begin{equation}
     g_m=\langle O^{*}_m({\bf s}) E_{\rm loc}({\bf s}) \rangle- \langle O^{*}_m({\bf s}) E_{\rm loc}({\bf s})\rangle, 
     \label{eq:gradient}
\end{equation}
where $\langle ... \rangle$ means the Monte Carlo average, and $O^{*}_m({\bf s})=\frac{1}{\Psi^{*}({\bf s})}\frac{\partial\Psi^{*}({\bf s})}{\partial \alpha^{*}_m}$. This enables efficient wavefunction optimization via stochastic gradient descent or stochastic reconfiguration. Detailed implementations are discussed in Refs.~\cite{liu2017,liu2021,liu2025}.

A critical feature of VMC arises from the {\it zero energy variance principle} for ground states. For energy eigenstates, i.e. $H\ket{\Psi}=E_g\ket{\Psi}$, the energy variance ${\rm var} \braket{H} = \braket{H^2}-\braket{H}^2=0$. In the context of Monte Carlo sampling, for energy eigenstates it has $E_{\rm loc}({\bf s})={\bra{{\bf s}}H|\Psi\rangle}/{\bra{{\bf s}}\Psi\rangle}=E_g$, which is independent of configurations $\ket {\bf s}$. Near the ground state, this property allows accurate energy estimation with minimal sampling noise, even using small sample sizes~\cite{sorella2017}. This is indeed what we observe, for example, in the comparison between PEPS and QMC results for the pure $\Z_2$ LGT presented in the main text. 

\section{S-2. VMC for gauge-invariant PEPS}

Below we outline key considerations for implementing variational Monte Carlo (VMC) with gauge-invariant PEPS, focusing on amplitude computation and Markov chain configuration sampling.

{\it Amplitude computations.} The gauge-invariant PEPS $\ket{\Psi}$ inherently eliminates unphysical configurations: Any configuration $\ket{\bf \bar s}$ violating the gauge symmetry has $\bra{{\bf \bar s}}\Psi\rangle=0$, enforced by the block-sparse structure of PEPS tensors. For physical configurations  $\ket{\vec s}\equiv \ket{\vec n, \vec p}$, considering the gauge canonical form,  the gauge constraint requires 
\begin{equation}
   {\rm mod} (p+n_l+n_d-n_u-n_r,N)=Q_{\rm x} 
\end{equation}
at each vertex $\vec x$ for $\Z_N$ gauge group, where $p$ denotes matter configuration, and $n_l, n_u, n_r, n_d$ represent gauge configurations on adjacent links..

The amplitude $\bra{{\bf s}}\Psi\rangle$ for a physical configuration $\ket{\vec s}\equiv \ket{\vec n, \vec p}$ corresponds to a tensor network:
\begin{equation}
\vcenter{\hbox{\includegraphics[scale=0.75, width=0.40\textwidth]{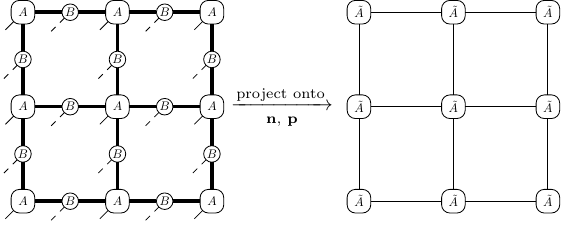}}},
\label{eq:project}
\end{equation}
where gauge symmetry reduces each tensor $A$ to a single sector $\tilde A$, with $B$ absent due to gauge canonical form. This reduces the bond dimension of amplitude networks to $D_k$, compared to the full PEPS bond dimension $D=\sum_k D_k$. The amplitude network are efficiently contracted using standard methods like SVD or variational compression~\cite{verstraete2008}.

{\it Configuration sampling.}  The sampling starts by randomly selecting an initial configuration $\ket{{\bf s}_{0}}$ that satisfy the gauge constraint: For every vertex $\vec x$ in the $\Z_N$ gauge theory, the condition ${\rm mod} (p+n_l+n_d-n_u-n_r,N)=Q_{\vec x}$ must hold. To generate a trial gauge configuration $\ket{{\bf s}_{1}}$, the plaquette operator $P_{\rm x}=U_{\vec x, 1} U_{\vec x+\vec e_1,2} U^\dag_{\vec x+\vec e_2,1}U^\dag_{\vec x,2}$ is applied to $\ket{{\bf s}_{0}}$, yielding $\ket{{\bf s}_{1}}=P^t_{\rm x} \ket{{\bf s}_{0}}$, where $t$ is an random integer in $[1,N-1]$ for $\Z_N$ gauge group. This procedure inherently preserves the gauge constraint, ensuring  $\ket{{\bf s}_{1}}$ remains physical. The Metropolis algorithm (as described previously) then determines whether $\ket{{\bf s}_{1}}$ is accepted into the Markov chain. For systems with dynamic matter fields, local updates are additionally performed on both gauge and matter configurations living on links, following $\ket{{\bf s}_{1}}=c_{\vec x}^\dagger U_{(\vec x, \alpha)} c_{\vec x + \vec e_\alpha} \ket{{\bf s}_{0}}$.

To accelerate sampling, all plaquettes and links are updated sequentially rather than randomly. This approach significantly reduces computational costs from $O(N^2_{\rm site})$ to $O(N_{\rm site})$~\cite{liu2021}. In pure gauge systems, updates focus exclusively on plaquettes, only involving gauge configurations. When matter fields are present, horizontal and vertical links are sequentially visited to update matter and gauge configurations, after each plaquette sweep over the lattice. After completing a full lattice sweep for plaquettes and links (termed a Monte Carlo sweep), physical observables are measured using the current configuration~\cite{liu2021}.

\change{
\section{S-3. time-dependent variational Monte Carlo for PEPS}

In (1+1)D, real-time evolution methods have been successfully established using MPS within frameworks including iTEBD~\cite{iTEBD} and the time-dependent variational principle (TDVP)~\cite{TDVP,TDVP2}. However, in (2+1)D, the intrinsic complexity of PEPS has hindered the development of corresponding real-time evolution methods, which remains in its early stages. Therefore, in this work, we employ the time-dependent variational Monte Carlo (t-VMC) approach~\cite{tvmc1,tvmc2,tvmc3}, which, to our knowledge, is applied here for the first time to PEPS. This method implements the TDVP and VMC  for variational wave functions to compute real-time evolution. It can be understood as applying the stochastic reconfiguration method~\cite{SR1998} to real time instead of imaginary time. The core step is to  solve a system of linear equations:
\begin{equation}
    iS_{mn}\dot{\alpha}_n=g_m,
    \label{eq:tVMC}
\end{equation}
where $S_{mn}$ is the overlap matrix:
\begin{equation}
S_{mn}=\langle O^{*}_m({\bf s})O_n({\bf s})\rangle-\langle O^{*}_m({\bf s})\rangle \langle O_n({\bf s})\rangle.
\end{equation}
The solution $\dot{\alpha}_m$ of the equation system is solved by iterative conjugate gradient method. When updating the variational parameters according to the ordinary differential equation (ODE) $\frac{d }{dt}\alpha_m=\dot{\alpha}_m$, we find a first-order update by $\alpha_m(t+\Delta t)= \alpha_m(t) - \Delta t*\dot{\alpha}_m(t)$ does not work well for real-time dynamics. To overcome this problem, one can use the fourth-order Runge-Kutta method to solve the ODE. Alternatively, we use a convenient method recently proposed in Ref.~\cite{nqstvmc_8} using Taylor expansion, sequentially updating the variational parameters twice in a time order per time step $\Delta t$ according to
\begin{align}
   & \alpha_m(t_1)= \alpha_m(t) - c_1*\Delta t*\dot{\alpha}_m(t),  \nonumber  \\
   & \alpha_m(t+\Delta t)= \alpha_m(t_1) - c_2*\Delta t*\dot{\alpha}_m(t_1), 
   \label{eq:secondorder}
\end{align}
where $c_{1,2}=(1\pm i)/2$. This gives a error $O(\Delta t^3)$ for expansions~\cite{nqstvmc_8}. Here we use $\Delta t=0.005$, which works very well in all cases. 
For a total time $T=18$, this needs 7200 evolution steps following Eqs.(\ref{eq:secondorder}). 
}

\section{S-4. Additional $\Z_N$ LGT results. }

\change{ 
We use $D_k=2$ which we find is good enough for convergence on relevant sizes (see Table \ref{tab:qmc_peps}) resulting in total PEPS bond dimension $D=4$, $6$, $8$, respectively for $\Z_2$, $\Z_3$ and $\Z_4$ LGTs. 

The $\Z_2$ LGT can be simulated unbiasedly by quantum Monte Carlo (QMC) via duality to the transverse field Ising model. Shown in Table \ref{tab:qmc_peps}, near the critical point $g_c\simeq0.3285$~\cite{wu2012phase}, PEPS energies agree excellently with QMC, indicating $D_k=2$ well converges the results.
Here QMC has slightly larger uncertainties due to critical slowing-down, whereas wavefunction-based PEPS show minimal sampling uncertainties due to the zero energy variance principle for ground state, as mentioned above.
}

\begin{figure}[tbp]
 \centering
\includegraphics[width=\textwidth]{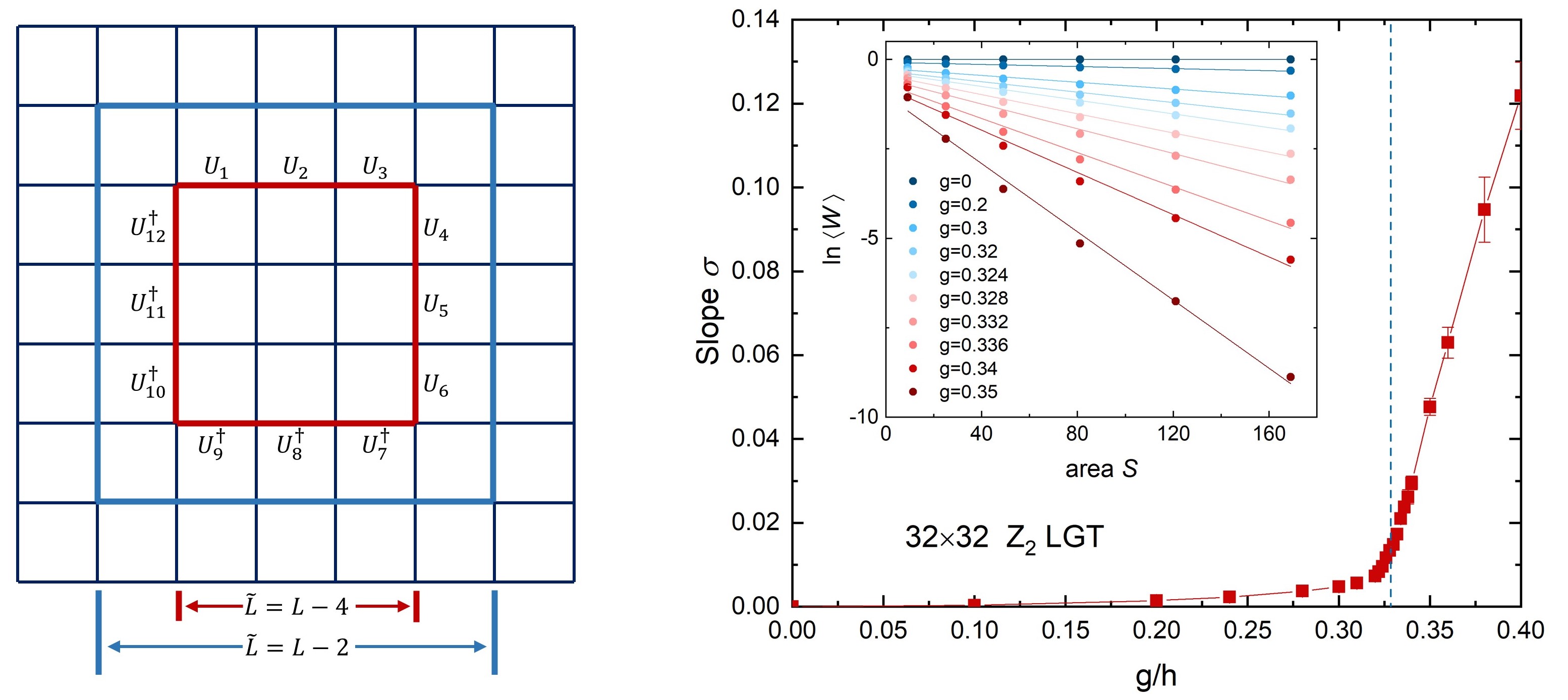}
 \caption{Left: Computing Wilson loop operators around a central square ${\tilde L} \times {\tilde L} $ on a given open boundary square lattice $L\times L$. Two different closed paths (red and blue) are highlighted, with  corresponding area $S=9$ and $S=25$. Right: The behavior of Wilson loop operator for $\Z_2$ LGT on a $32\times 32$ lattice. The inset shows the linear fits of $\ln \braket{W} \propto {-\sigma S}$  to extracted the slope $\sigma$; the main panel shows the variation of $\sigma$ with $g/h$ increasing, and the vertical bule dashed line denotes the critical point $g_c\simeq 0.3285$ from quantum Monte Carlo~\cite{wu2012phase}.     }
 \label{fig:square}
 \end{figure}

\setlength{\tabcolsep}{12pt}
\begin{table}
\centering
\caption{Ground state energy per site. The top part shows QMC and GI-PEPS ($D=4$) energy comparison for $16\times16$ $\Z_2$ LGT.
The rest shows the $D$-convergence of GI-PEPS energy at critical points for the largest sizes, i.e. $24\times 24$ $\Z_3$ LGT and $20\times 20$ $\Z_4$ LGT, respectively.}
\label{tab:qmc_peps}
\begin{tabular}{ccc}
\toprule
\textbf{$\Z_2$} & \textbf{QMC} & \textbf{PEPS} \\
$g=0.30$  & $-0.76400(3)$ & $-0.763973(8)$ \\
$g=0.31$ & $-0.73841(6)$ & $-0.738443(6)$ \\
$g=0.32$ & $-0.71400(7)$ & $-0.714032(8)$ \\
$g=0.33$ & $-0.69091(7)$  & $-0.690923(6)$ \\
$g=0.34$ & $-0.6691(1)$  & $-0.669161(6)$ \\
$g=0.35$ & $-0.6486(1)$  & $-0.648714(6)$ \\ 
\midrule
\textbf{$\Z_3$} &  $D=6$ &  $D=9$ \\
$g=0.375$ & $-0.548401(6)$  & $-0.548409(4)$ \\ 
\midrule
\textbf{$\Z_4$} & $D=8$ & $D=12$ \\
$g=0.33$ & $-0.712554(8)$  & $-0.712557(4)$ \\ 
\bottomrule
\end{tabular}
\end{table}

{\it Wislon loop operator.} The Wilson loop operator \( W \) is evaluated along a closed square path of dimensions \( \tilde{L} \times \tilde{L} \). As illustrated in the left panel of Fig.~\ref{fig:square} for a \( 3 \times 3 \) lattice (red lines), this operator takes the form 
\[
W = U_1 \otimes \cdots \otimes U_6 \otimes U^{\dagger}_7 \otimes \cdots \otimes U^{\dagger}_{12}, 
\]
acting on the gauge field variables along the closed contour. Its expectation value \( \langle W \rangle \) can be efficiently computed via Monte Carlo sampling. For an \( L \times L \) lattice, we select a series of concentric \( \tilde{L} \times \tilde{L} \) closed paths and calculate \( \langle W \rangle \) for each path, where \( S \) denotes the area enclosed by the loop.

In pure \( \mathbb{Z}_2 \) lattice gauge theory, the Wilson loop exhibits distinct scaling behaviors across phases: perimeter-law scaling in the deconfined phase and area-law scaling in the confined phase. This transition can be quantified through the string tension \( \sigma \), obtained from the scaling relation \( \langle W \rangle \propto e^{-\sigma S}. \)
We determine \( \sigma \) by performing linear fittins on \( \ln\langle W \rangle \) versus \( S \), as shown in the right panel of Fig.~\ref{fig:square} based on the $32\times 32$ lattice. The evolution of \( \sigma \) with \( g/h \) clearly demonstrates a phase transition between the deconfined regime (small \( \sigma \)) and confined regime (large \( \sigma \)).

In addition, for \( \mathbb{Z}_2 \) gauge theory coupled to hard-core bosons, in the main text we employ different bulk region definitions to estimate thermodynamic-limit energies~\cite{liu2021}. Specifically, the blue contour in Fig.~\ref{fig:square}(left) demarcates a central \( (L-2) \times (L-2) \) region, while the red contour corresponds to a \( (L-4) \times (L-4) \) region. By analyzing these progressively smaller bulk regions across varying lattice sizes \( L \times L \), we perform systematic finite-size extrapolations, as shown in the main text.

\begin{figure}[htb]
 \centering
  \includegraphics[width=0.6\textwidth]{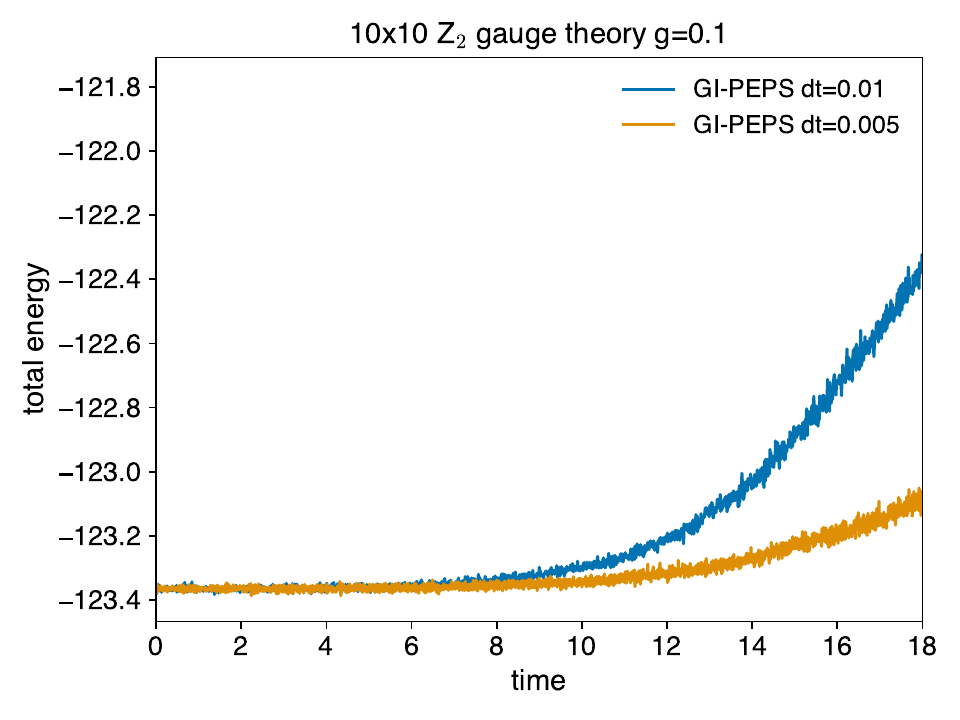}
  \caption{Energy conservation of the vison real-time dynamics on a $10\times 10$ $\Z_2$ gauge theory.
  The bond dimension of the GI-PEPS is 8.}
  \label{fig:vison_energy}
\end{figure}

\begin{figure}[htb]
 \centering
\includegraphics[width=0.6\textwidth]{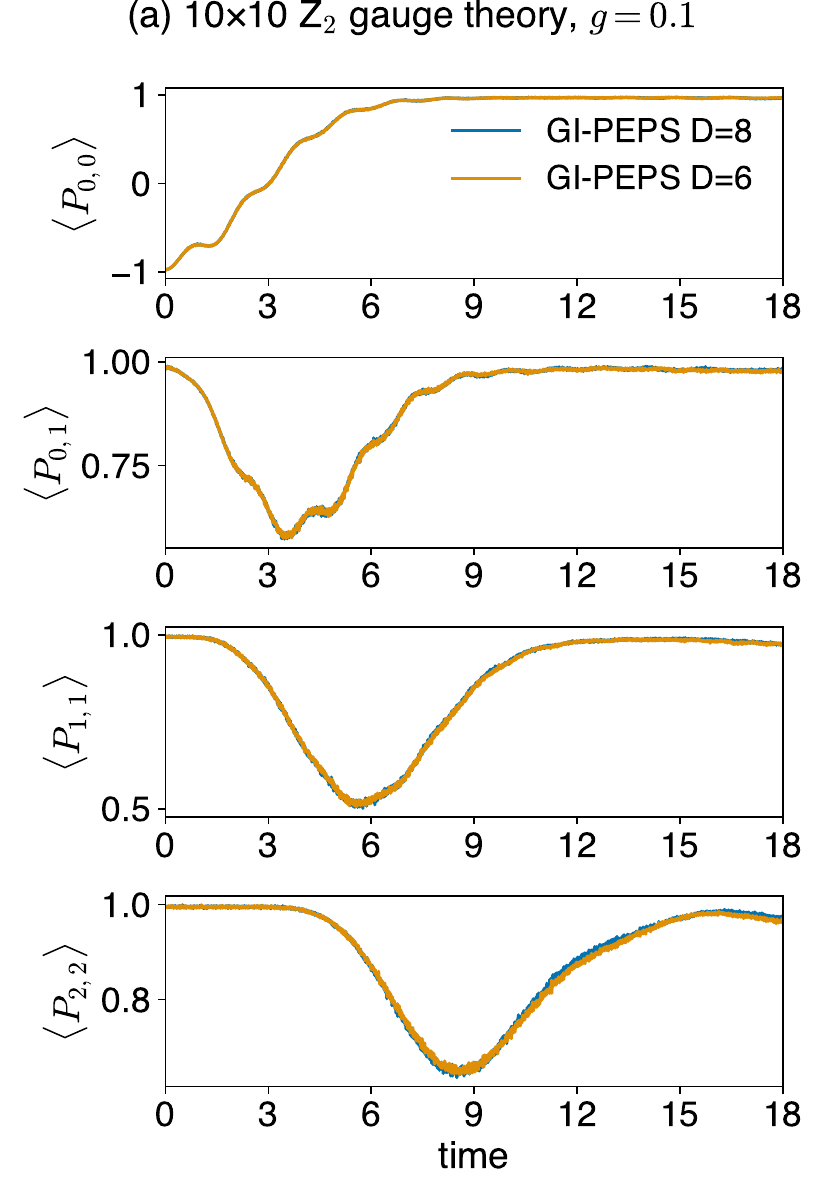}
  \caption{Selected plaquette values of the vison real-time dynamics on a $10\times 10$ $\Z_2$ gauge theory at bond dimension 6 and 8.
  $\Delta t = 0.005$ for both bond dimensions.}
  \label{fig:vison_D6_D8}
\end{figure}

\change{

\section{S-5. Comparison with iPEPS results for $\Z_3$ LGT}

The ground states of the $\mathbb{Z}_3$ lattice gauge theory (LGT) have been studied using the iPEPS method with a non-gauge-constrained ansatz, optimized via simple update imaginary time evolution. To benchmark our results, we compare the thermodynamic limit ground state energy obtained from our finite-size extrapolations with the corresponding iPEPS energies reported in Ref.~\cite{iPEPS_LGT2021}.

Note that the Hamiltonian parameter $g$ used in this work corresponds to the coupling parameter $\tilde{g}^2$ in Ref.~\cite{iPEPS_LGT2021} via the relation $g = {(\tilde{g}^2)}^2 / 3$ (set $h=1$). To facilitate comparison, we extrapolate our finite-size energies to the thermodynamic limit. The inset of Fig.~\ref{fig:ipeps} illustrates this extrapolation procedure for the representative case $g=0.35$. The second and third-order polynomial fits give almost same extrapolated values.

The main panel of Fig.~\ref{fig:ipeps} presents both the non-gauge-constrained iPEPS energies and our extrapolated thermodynamic limit energies across a range of $g$ values. At small $g$, the agreement is excellent. However, for $g \geq 0.35$, our energies become lower (as exemplified for $g=0.35$ in the inset) and exhibit a smooth variation with $g$. Analysis of the first and second derivatives of our energy curve indicates a phase transition at $g_c = 0.375(3)$ (see main text). This agrees well with a recent estimate of $g_c = 0.37(1)$ obtained using neural quantum states on tori up to size $10\times 10$~\cite{Apte2024deep}. In contrast, the iPEPS energies exhibit non-smooth behavior in this critical region and yield an estimated critical point at $g_c=0.448(3)$~\cite{iPEPS_LGT2021}. This discrepancy likely arises from insufficient iPEPS accuracy near the critical point.

\begin{figure}[htb]
 \centering
  \includegraphics[scale=0.5]{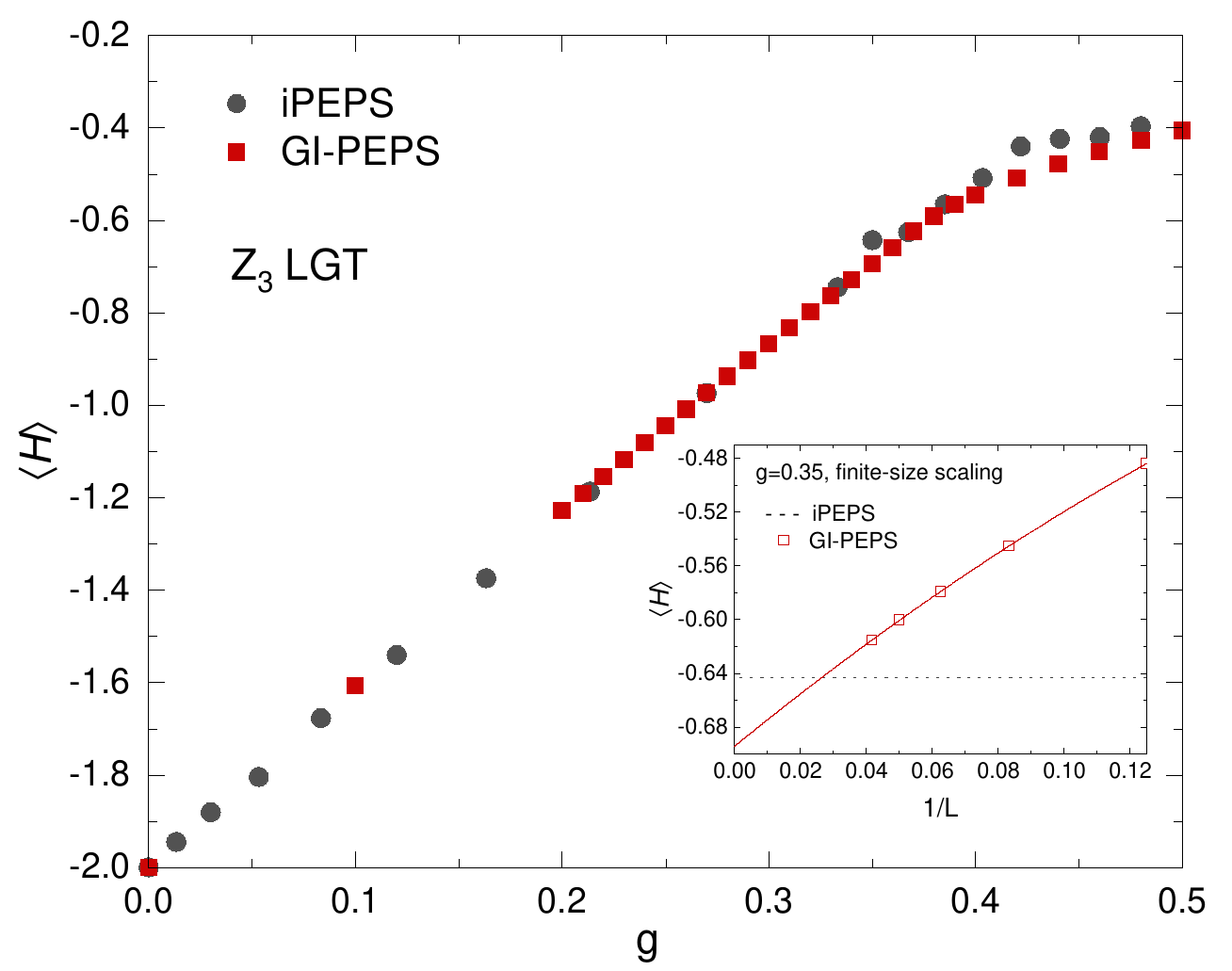}
  \caption{Comparison of $\Z_3$ LGT ground state energies ($h=1$). The inset shows the finite-size scaling of our GI-PEPS results on $L\times L$ with a second-order polynomial extrapolation including $L=8, 12, 16, 20, 24$, and the black-dashed line denotes the iPEPS energy, at $g=0.35$. The main panel shows the thermodynamic limit energies obtained from extrapolations of our GI-PEPS energies (extrapolation errors are about $0.0001$), compared with iPEPS results taken from Ref.~\cite{iPEPS_LGT2021}, at various $g$.}
  \label{fig:ipeps}
\end{figure}

\section{S-6. Additional data to real-time dynamics of the 10$\times$10 $\Z_2$ gauge theory}
In Fig. \ref{fig:vison_energy},  we present the total energy of the system during the real-time simulation described in the main text for the $10\times 10$ vison dynamics in a $\Z_2$ gauge theory.
We use two time steps $\Delta t= 0.01$ and $0.005$. 
We see that for $\Delta t = 0.005$ (which is adopted in the main text), the energy drift is less than $0.2\%$ during the entire time trajectory. 

In Fig. \ref{fig:vison_D6_D8}, we present the selected plaquette values at bond dimension $D=6$ and $D=8$ to demonstrate the convergence with respect to the bond dimension.
}

\end{document}

%% file: tikz/cdagc_AA.tex
\begin{tikzpicture}[baseline = (X.base),every node/.style={scale=0.6},scale=.7]
\draw (0.5,0) -- (1,0);
\draw[rounded corners] (1,0.5) rectangle (2,-0.5);
\draw (1.5,0) node [scale=1.1]{$A'_x$};   \draw (1.5,-.5) -- (1.5,-1.0);
\draw (2,0) -- (2.5,0); 
\draw[rounded corners] (2.5,0.5) rectangle (3.5,-0.5);
\draw (3.0,0) node [scale=1.1] {$A'_{x+1}$};       \draw (3,-.5) -- (3,-1.0); 
\draw (3.5,0) -- (4.0,0);
\draw (2,0) node(X) {$\phantom{X}$};
\end{tikzpicture}  
\equiv
\begin{tikzpicture}[baseline = (X.base),every node/.style={scale=0.6},scale=.7]
\draw (0.5,0) -- (1,0);
\draw[rounded corners] (1,0.5) rectangle (2,-0.5);
\draw (1.5,0) node [scale=1.1]{$A_x$};   \draw (1.5,-.5) -- (1.5,-1.0);
\draw (2,0) -- (2.5,0); 
\draw[rounded corners] (2.5,0.5) rectangle (3.5,-0.5);
\draw (3.0,0) node [scale=1.1] {$A_{x+1}$};       \draw (3,-.5) -- (3,-1.0); 
\draw (3.5,0) -- (4.0,0);
\draw[rounded corners] (1.,-1.0) rectangle (3.5,-1.75); 
\draw (2.25,-1.375) node [scale=1.1] {$c_x^\dag c_{x+1}$};  \draw (1.5,-1.75) -- (1.5,-2.35);
\draw (3.0,-1.75) -- (3.0,-2.35);
\draw (2,0) node(X) {$\phantom{X}$};
\end{tikzpicture}  

%% file: tikz/cdagc_AA_noLGT.tex
\begin{tikzpicture}[baseline = (X.base),every node/.style={scale=0.6},scale=.7]
\draw (0.5,0) -- (1,0);
\draw[rounded corners] (1,0.5) rectangle (2,-0.5);
\draw (1.5,0) node [scale=1.1]{$A_x$};   \draw (1.5,-.5) -- (1.5,-1.0);
\draw (2,0) -- (2.5,0); 
\draw (2.85,0) circle (0.35);  
\draw (2.85,0) node [scale=1.1]{$B$};       \draw[dashed] (2.85,-.35) -- (2.85,-1.0);
\draw (3.2,0) -- (3.7,0);
\draw[rounded corners] (3.7,0.5) rectangle (4.7,-0.5);
\draw (4.2,0) node [scale=1.1] {$A_{x+1}$};       \draw (4.2,-.5) -- (4.2,-1.0); 
\draw (4.7,0) -- (5.2,0);
\draw[rounded corners] (1.,-1.0) rectangle (4.7,-1.75); 
\draw (2.85,-1.375) node [scale=1.1] {$c_x^\dag U_{(x,1)} c_{x+1}$};  
\draw (1.5,-1.75) -- (1.5,-2.35);
\draw[dashed] (2.85,-1.75) -- (2.85,-2.35);
\draw (4.2,-1.75) -- (4.2,-2.35);
\draw (5,0) node(X) {$\phantom{X}$};
\end{tikzpicture}  
=
\begin{tikzpicture}[baseline = (X.base),every node/.style={scale=0.6},scale=.7]
\draw (0.5,0) -- (1,0);
\draw[rounded corners] (1,0.5) rectangle (2,-0.5);
\draw (1.5,0) node [scale=1.1]{$A'_x$};   \draw (1.5,-.5) -- (1.5,-1.0);
\draw (2,0) -- (2.5,0); 
\draw (2.85,0) circle (0.35);  
\draw (2.85,0) node [scale=1.1]{$B$};       \draw[dashed] (2.85,-.35) -- (2.85,-1.0);
\draw (3.2,0) -- (3.7,0);
\draw[rounded corners] (3.7,0.5) rectangle (4.7,-0.5);
\draw (4.2,0) node [scale=1.1] {$A'_{x+1}$};       \draw (4.2,-.5) -- (4.2,-1.0); 
\draw (4.7,0) -- (5.2,0);
\draw (5,0) node (X) {$\phantom{X}$};
\end{tikzpicture}  

%% file: tikz/rho_MPS.tex
\begin{tikzpicture}[baseline = (X.base),every node/.style={scale=0.6},scale=.7]
\draw[rounded corners] (0,0.5) rectangle (0.9,1.4);
\draw (0.45,0.95) node [scale=1.1]{$A$};   \draw (0.9,0.95) -- (1.4,.95);
\draw (1.8,0.95) circle (0.4); \draw (2.2,0.95) -- (2.6,.95);
\draw (1.8,0.95) node [scale=1.0]{$B$};
\draw [dashed] (1.8,0.55) -- (1.8, -.55);
\draw (0.5,-.5) -- (0.5,.5);
\draw[rounded corners] (0,-0.5) rectangle (0.9,-1.4);
\draw (0.45,-0.95) node [scale=1.1]{$A^*$};\draw (0.9,-0.95) -- (1.4,-.95);
\draw (1.8,-0.95) circle (0.4); \draw (2.2,-0.95) -- (2.6,-.95);
\draw (1.8,-0.95) node [scale=1.0]{$B^*$};
\draw (-0.5,0.) circle (0.3);  \draw (-0.5,0.) node [scale=1.1]{$\rho$};
\draw (-0.5,0.3) edge[in=180,out=90] (0,0.9);
\draw (-0.5,-0.3) edge[in=180,out=-90] (0,-0.9);
\draw (0,0) node(X) {$\phantom{X}$}; 
\end{tikzpicture}  
=
\begin{tikzpicture}[baseline = (X.base),every node/.style={scale=0.6},scale=.7]
\draw[rounded corners] (0,0.5) rectangle (0.9,1.4);
\draw (0.45,0.95) node [scale=1.1]{$A$};  \draw (0.9,0.95) -- (1.4,.95);
\draw (0.5,-.5) -- (0.5,.5); 
\draw[rounded corners] (0,-0.5) rectangle (0.9,-1.4);
\draw (0.45,-0.95) node [scale=1.1]{$A^*$}; \draw (0.9,-0.95) -- (1.4,-.95);
\draw (-0.5,0.) circle (0.3);  \draw (-0.5,0.) node [scale=1.1]{$\rho$};
\draw (-0.5,0.3) edge[in=180,out=90] (0,0.9);
\draw (-0.5,-0.3) edge[in=180,out=-90] (0,-0.9);
\draw (0,0) node(X) {$\phantom{X}$}; 
\end{tikzpicture}

%% file: main_v2.bbl
\begin{thebibliography}{93}%
\makeatletter
\providecommand \@ifxundefined [1]{%
 \@ifx{#1\undefined}
}%
\providecommand \@ifnum [1]{%
 \ifnum #1\expandafter \@firstoftwo
 \else \expandafter \@secondoftwo
 \fi
}%
\providecommand \@ifx [1]{%
 \ifx #1\expandafter \@firstoftwo
 \else \expandafter \@secondoftwo
 \fi
}%
\providecommand \natexlab [1]{#1}%
\providecommand \enquote  [1]{``#1''}%
\providecommand \bibnamefont  [1]{#1}%
\providecommand \bibfnamefont [1]{#1}%
\providecommand \citenamefont [1]{#1}%
\providecommand \href@noop [0]{\@secondoftwo}%
\providecommand \href [0]{\begingroup \@sanitize@url \@href}%
\providecommand \@href[1]{\@@startlink{#1}\@@href}%
\providecommand \@@href[1]{\endgroup#1\@@endlink}%
\providecommand \@sanitize@url [0]{\catcode `\\12\catcode `\$12\catcode
  `\&12\catcode `\#12\catcode `\^12\catcode `\_12\catcode `\%12\relax}%
\providecommand \@@startlink[1]{}%
\providecommand \@@endlink[0]{}%
\providecommand \url  [0]{\begingroup\@sanitize@url \@url }%
\providecommand \@url [1]{\endgroup\@href {#1}{\urlprefix }}%
\providecommand \urlprefix  [0]{URL }%
\providecommand \Eprint [0]{\href }%
\providecommand \doibase [0]{https://doi.org/}%
\providecommand \selectlanguage [0]{\@gobble}%
\providecommand \bibinfo  [0]{\@secondoftwo}%
\providecommand \bibfield  [0]{\@secondoftwo}%
\providecommand \translation [1]{[#1]}%
\providecommand \BibitemOpen [0]{}%
\providecommand \bibitemStop [0]{}%
\providecommand \bibitemNoStop [0]{.\EOS\space}%
\providecommand \EOS [0]{\spacefactor3000\relax}%
\providecommand \BibitemShut  [1]{\csname bibitem#1\endcsname}%
\let\auto@bib@innerbib\@empty
\bibitem [{\citenamefont {Wilson}(1974)}]{wilson1974}%
  \BibitemOpen
  \bibfield  {author} {\bibinfo {author} {\bibfnamefont {K.~G.}\ \bibnamefont
  {Wilson}},\ }\bibfield  {title} {\bibinfo {title} {Confinement of quarks},\
  }\href {https://doi.org/10.1103/PhysRevD.10.2445} {\bibfield  {journal}
  {\bibinfo  {journal} {Phys. Rev. D}\ }\textbf {\bibinfo {volume} {10}},\
  \bibinfo {pages} {2445} (\bibinfo {year} {1974})}\BibitemShut {NoStop}%
\bibitem [{\citenamefont {Kogut}(1979)}]{kogut1979}%
  \BibitemOpen
  \bibfield  {author} {\bibinfo {author} {\bibfnamefont {J.~B.}\ \bibnamefont
  {Kogut}},\ }\bibfield  {title} {\bibinfo {title} {An introduction to lattice
  gauge theory and spin systems},\ }\href
  {https://doi.org/10.1103/RevModPhys.51.659} {\bibfield  {journal} {\bibinfo
  {journal} {Rev. Mod. Phys.}\ }\textbf {\bibinfo {volume} {51}},\ \bibinfo
  {pages} {659} (\bibinfo {year} {1979})}\BibitemShut {NoStop}%
\bibitem [{\citenamefont {Fukushima}\ and\ \citenamefont
  {Hatsuda}(2010)}]{Fukushima2011}%
  \BibitemOpen
  \bibfield  {author} {\bibinfo {author} {\bibfnamefont {K.}~\bibnamefont
  {Fukushima}}\ and\ \bibinfo {author} {\bibfnamefont {T.}~\bibnamefont
  {Hatsuda}},\ }\bibfield  {title} {\bibinfo {title} {The phase diagram of
  dense qcd},\ }\href {https://doi.org/10.1088/0034-4885/74/1/014001}
  {\bibfield  {journal} {\bibinfo  {journal} {Reports on Progress in Physics}\
  }\textbf {\bibinfo {volume} {74}},\ \bibinfo {pages} {014001} (\bibinfo
  {year} {2010})}\BibitemShut {NoStop}%
\bibitem [{\citenamefont {Wen}(2004)}]{wen2004quantum}%
  \BibitemOpen
  \bibfield  {author} {\bibinfo {author} {\bibfnamefont {X.-G.}\ \bibnamefont
  {Wen}},\ }\href@noop {} {\emph {\bibinfo {title} {Quantum Field Theory of
  Many Body Systems: From the origin of sound to an origin of light and
  electrons}}}\ (\bibinfo  {publisher} {Oxford University Press, New York},\
  \bibinfo {year} {2004})\BibitemShut {NoStop}%
\bibitem [{\citenamefont {Kitaev}(2006)}]{kitaev2006anyons}%
  \BibitemOpen
  \bibfield  {author} {\bibinfo {author} {\bibfnamefont {A.}~\bibnamefont
  {Kitaev}},\ }\bibfield  {title} {\bibinfo {title} {Anyons in an exactly
  solved model and beyond},\ }\href
  {https://doi.org/https://doi.org/10.1016/j.aop.2005.10.005} {\bibfield
  {journal} {\bibinfo  {journal} {Annals of Physics}\ }\textbf {\bibinfo
  {volume} {321}},\ \bibinfo {pages} {2} (\bibinfo {year} {2006})}\BibitemShut
  {NoStop}%
\bibitem [{\citenamefont {Troyer}\ and\ \citenamefont
  {Wiese}(2005)}]{troyer2005}%
  \BibitemOpen
  \bibfield  {author} {\bibinfo {author} {\bibfnamefont {M.}~\bibnamefont
  {Troyer}}\ and\ \bibinfo {author} {\bibfnamefont {U.-J.}\ \bibnamefont
  {Wiese}},\ }\bibfield  {title} {\bibinfo {title} {Computational complexity
  and fundamental limitations to fermionic quantum monte carlo simulations},\
  }\href {https://doi.org/10.1103/PhysRevLett.94.170201} {\bibfield  {journal}
  {\bibinfo  {journal} {Phys. Rev. Lett.}\ }\textbf {\bibinfo {volume} {94}},\
  \bibinfo {pages} {170201} (\bibinfo {year} {2005})}\BibitemShut {NoStop}%
\bibitem [{\citenamefont {Gross}\ and\ \citenamefont
  {Bloch}(2017)}]{gross2017quantum}%
  \BibitemOpen
  \bibfield  {author} {\bibinfo {author} {\bibfnamefont {C.}~\bibnamefont
  {Gross}}\ and\ \bibinfo {author} {\bibfnamefont {I.}~\bibnamefont {Bloch}},\
  }\bibfield  {title} {\bibinfo {title} {Quantum simulations with ultracold
  atoms in optical lattices},\ }\href {https://doi.org/10.1126/science.aal3837}
  {\bibfield  {journal} {\bibinfo  {journal} {Science}\ }\textbf {\bibinfo
  {volume} {357}},\ \bibinfo {pages} {995} (\bibinfo {year}
  {2017})}\BibitemShut {NoStop}%
\bibitem [{\citenamefont {Bruzewicz}\ \emph {et~al.}(2019)\citenamefont
  {Bruzewicz}, \citenamefont {Chiaverini}, \citenamefont {McConnell},\ and\
  \citenamefont {Sage}}]{bruzewicz2019trapped}%
  \BibitemOpen
  \bibfield  {author} {\bibinfo {author} {\bibfnamefont {C.~D.}\ \bibnamefont
  {Bruzewicz}}, \bibinfo {author} {\bibfnamefont {J.}~\bibnamefont
  {Chiaverini}}, \bibinfo {author} {\bibfnamefont {R.}~\bibnamefont
  {McConnell}},\ and\ \bibinfo {author} {\bibfnamefont {J.~M.}\ \bibnamefont
  {Sage}},\ }\bibfield  {title} {\bibinfo {title} {Trapped-ion quantum
  computing: Progress and challenges},\ }\href
  {https://doi.org/10.1063/1.5088164} {\bibfield  {journal} {\bibinfo
  {journal} {Applied Physics Reviews}\ }\textbf {\bibinfo {volume} {6}}
  (\bibinfo {year} {2019})}\BibitemShut {NoStop}%
\bibitem [{\citenamefont {Krantz}\ \emph {et~al.}(2019)\citenamefont {Krantz},
  \citenamefont {Kjaergaard}, \citenamefont {Yan}, \citenamefont {Orlando},
  \citenamefont {Gustavsson},\ and\ \citenamefont
  {Oliver}}]{krantz2019quantum}%
  \BibitemOpen
  \bibfield  {author} {\bibinfo {author} {\bibfnamefont {P.}~\bibnamefont
  {Krantz}}, \bibinfo {author} {\bibfnamefont {M.}~\bibnamefont {Kjaergaard}},
  \bibinfo {author} {\bibfnamefont {F.}~\bibnamefont {Yan}}, \bibinfo {author}
  {\bibfnamefont {T.~P.}\ \bibnamefont {Orlando}}, \bibinfo {author}
  {\bibfnamefont {S.}~\bibnamefont {Gustavsson}},\ and\ \bibinfo {author}
  {\bibfnamefont {W.~D.}\ \bibnamefont {Oliver}},\ }\bibfield  {title}
  {\bibinfo {title} {A quantum engineer's guide to superconducting qubits},\
  }\href {https://doi.org/10.1063/1.5089550} {\bibfield  {journal} {\bibinfo
  {journal} {Applied physics reviews}\ }\textbf {\bibinfo {volume} {6}}
  (\bibinfo {year} {2019})}\BibitemShut {NoStop}%
\bibitem [{\citenamefont {Yang}\ \emph {et~al.}(2020)\citenamefont {Yang},
  \citenamefont {Sun}, \citenamefont {Ott}, \citenamefont {Wang}, \citenamefont
  {Zache}, \citenamefont {Halimeh}, \citenamefont {Yuan}, \citenamefont
  {Hauke},\ and\ \citenamefont {Pan}}]{yang2020observation}%
  \BibitemOpen
  \bibfield  {author} {\bibinfo {author} {\bibfnamefont {B.}~\bibnamefont
  {Yang}}, \bibinfo {author} {\bibfnamefont {H.}~\bibnamefont {Sun}}, \bibinfo
  {author} {\bibfnamefont {R.}~\bibnamefont {Ott}}, \bibinfo {author}
  {\bibfnamefont {H.-Y.}\ \bibnamefont {Wang}}, \bibinfo {author}
  {\bibfnamefont {T.~V.}\ \bibnamefont {Zache}}, \bibinfo {author}
  {\bibfnamefont {J.~C.}\ \bibnamefont {Halimeh}}, \bibinfo {author}
  {\bibfnamefont {Z.-S.}\ \bibnamefont {Yuan}}, \bibinfo {author}
  {\bibfnamefont {P.}~\bibnamefont {Hauke}},\ and\ \bibinfo {author}
  {\bibfnamefont {J.-W.}\ \bibnamefont {Pan}},\ }\bibfield  {title} {\bibinfo
  {title} {Observation of gauge invariance in a 71-site \text{Bose-Hubbard}
  quantum simulator},\ }\href {https://doi.org/10.1038/s41586-020-2910-8}
  {\bibfield  {journal} {\bibinfo  {journal} {Nature}\ }\textbf {\bibinfo
  {volume} {587}},\ \bibinfo {pages} {392} (\bibinfo {year}
  {2020})}\BibitemShut {NoStop}%
\bibitem [{\citenamefont {Banuls}\ \emph {et~al.}(2020)\citenamefont {Banuls},
  \citenamefont {Blatt}, \citenamefont {Catani}, \citenamefont {Celi},
  \citenamefont {Cirac}, \citenamefont {Dalmonte}, \citenamefont {Fallani},
  \citenamefont {Jansen}, \citenamefont {Lewenstein}, \citenamefont
  {Montangero} \emph {et~al.}}]{banuls2020simulating}%
  \BibitemOpen
  \bibfield  {author} {\bibinfo {author} {\bibfnamefont {M.~C.}\ \bibnamefont
  {Banuls}}, \bibinfo {author} {\bibfnamefont {R.}~\bibnamefont {Blatt}},
  \bibinfo {author} {\bibfnamefont {J.}~\bibnamefont {Catani}}, \bibinfo
  {author} {\bibfnamefont {A.}~\bibnamefont {Celi}}, \bibinfo {author}
  {\bibfnamefont {J.~I.}\ \bibnamefont {Cirac}}, \bibinfo {author}
  {\bibfnamefont {M.}~\bibnamefont {Dalmonte}}, \bibinfo {author}
  {\bibfnamefont {L.}~\bibnamefont {Fallani}}, \bibinfo {author} {\bibfnamefont
  {K.}~\bibnamefont {Jansen}}, \bibinfo {author} {\bibfnamefont
  {M.}~\bibnamefont {Lewenstein}}, \bibinfo {author} {\bibfnamefont
  {S.}~\bibnamefont {Montangero}}, \emph {et~al.},\ }\bibfield  {title}
  {\bibinfo {title} {Simulating lattice gauge theories within quantum
  technologies},\ }\href {https://doi.org/10.1140/epjd/e2020-100571-8}
  {\bibfield  {journal} {\bibinfo  {journal} {The European physical journal D}\
  }\textbf {\bibinfo {volume} {74}},\ \bibinfo {pages} {1} (\bibinfo {year}
  {2020})}\BibitemShut {NoStop}%
\bibitem [{\citenamefont {Tagliacozzo}\ and\ \citenamefont
  {Vidal}(2011)}]{Tagliacozzo2011entangle}%
  \BibitemOpen
  \bibfield  {author} {\bibinfo {author} {\bibfnamefont {L.}~\bibnamefont
  {Tagliacozzo}}\ and\ \bibinfo {author} {\bibfnamefont {G.}~\bibnamefont
  {Vidal}},\ }\bibfield  {title} {\bibinfo {title} {Entanglement
  renormalization and gauge symmetry},\ }\href
  {https://doi.org/10.1103/PhysRevB.83.115127} {\bibfield  {journal} {\bibinfo
  {journal} {Phys. Rev. B}\ }\textbf {\bibinfo {volume} {83}},\ \bibinfo
  {pages} {115127} (\bibinfo {year} {2011})}\BibitemShut {NoStop}%
\bibitem [{\citenamefont {Tagliacozzo}\ \emph {et~al.}(2014)\citenamefont
  {Tagliacozzo}, \citenamefont {Celi},\ and\ \citenamefont
  {Lewenstein}}]{Tagliacozzo2014tensor}%
  \BibitemOpen
  \bibfield  {author} {\bibinfo {author} {\bibfnamefont {L.}~\bibnamefont
  {Tagliacozzo}}, \bibinfo {author} {\bibfnamefont {A.}~\bibnamefont {Celi}},\
  and\ \bibinfo {author} {\bibfnamefont {M.}~\bibnamefont {Lewenstein}},\
  }\bibfield  {title} {\bibinfo {title} {Tensor networks for lattice gauge
  theories with continuous groups},\ }\href
  {https://doi.org/10.1103/PhysRevX.4.041024} {\bibfield  {journal} {\bibinfo
  {journal} {Phys. Rev. X}\ }\textbf {\bibinfo {volume} {4}},\ \bibinfo {pages}
  {041024} (\bibinfo {year} {2014})}\BibitemShut {NoStop}%
\bibitem [{\citenamefont {Buyens}\ \emph {et~al.}(2014)\citenamefont {Buyens},
  \citenamefont {Haegeman}, \citenamefont {Van~Acoleyen}, \citenamefont
  {Verschelde},\ and\ \citenamefont {Verstraete}}]{buyens2014matrix}%
  \BibitemOpen
  \bibfield  {author} {\bibinfo {author} {\bibfnamefont {B.}~\bibnamefont
  {Buyens}}, \bibinfo {author} {\bibfnamefont {J.}~\bibnamefont {Haegeman}},
  \bibinfo {author} {\bibfnamefont {K.}~\bibnamefont {Van~Acoleyen}}, \bibinfo
  {author} {\bibfnamefont {H.}~\bibnamefont {Verschelde}},\ and\ \bibinfo
  {author} {\bibfnamefont {F.}~\bibnamefont {Verstraete}},\ }\bibfield  {title}
  {\bibinfo {title} {Matrix product states for gauge field theories},\ }\href
  {https://doi.org/10.1103/PhysRevLett.113.091601} {\bibfield  {journal}
  {\bibinfo  {journal} {Phys. Rev. Lett.}\ }\textbf {\bibinfo {volume} {113}},\
  \bibinfo {pages} {091601} (\bibinfo {year} {2014})}\BibitemShut {NoStop}%
\bibitem [{\citenamefont {Silvi}\ \emph {et~al.}(2014)\citenamefont {Silvi},
  \citenamefont {Rico}, \citenamefont {Calarco},\ and\ \citenamefont
  {Montangero}}]{Silvi2014}%
  \BibitemOpen
  \bibfield  {author} {\bibinfo {author} {\bibfnamefont {P.}~\bibnamefont
  {Silvi}}, \bibinfo {author} {\bibfnamefont {E.}~\bibnamefont {Rico}},
  \bibinfo {author} {\bibfnamefont {T.}~\bibnamefont {Calarco}},\ and\ \bibinfo
  {author} {\bibfnamefont {S.}~\bibnamefont {Montangero}},\ }\bibfield  {title}
  {\bibinfo {title} {Lattice gauge tensor networks},\ }\href
  {https://doi.org/10.1088/1367-2630/16/10/103015} {\bibfield  {journal}
  {\bibinfo  {journal} {New Journal of Physics}\ }\textbf {\bibinfo {volume}
  {16}},\ \bibinfo {pages} {103015} (\bibinfo {year} {2014})}\BibitemShut
  {NoStop}%
\bibitem [{\citenamefont {Zou}\ \emph {et~al.}(2014)\citenamefont {Zou},
  \citenamefont {Liu}, \citenamefont {Lai}, \citenamefont {Unmuth-Yockey},
  \citenamefont {Yang}, \citenamefont {Bazavov}, \citenamefont {Xie},
  \citenamefont {Xiang}, \citenamefont {Chandrasekharan}, \citenamefont
  {Tsai},\ and\ \citenamefont {Meurice}}]{zou2014}%
  \BibitemOpen
  \bibfield  {author} {\bibinfo {author} {\bibfnamefont {H.}~\bibnamefont
  {Zou}}, \bibinfo {author} {\bibfnamefont {Y.}~\bibnamefont {Liu}}, \bibinfo
  {author} {\bibfnamefont {C.-Y.}\ \bibnamefont {Lai}}, \bibinfo {author}
  {\bibfnamefont {J.}~\bibnamefont {Unmuth-Yockey}}, \bibinfo {author}
  {\bibfnamefont {L.-P.}\ \bibnamefont {Yang}}, \bibinfo {author}
  {\bibfnamefont {A.}~\bibnamefont {Bazavov}}, \bibinfo {author} {\bibfnamefont
  {Z.~Y.}\ \bibnamefont {Xie}}, \bibinfo {author} {\bibfnamefont
  {T.}~\bibnamefont {Xiang}}, \bibinfo {author} {\bibfnamefont
  {S.}~\bibnamefont {Chandrasekharan}}, \bibinfo {author} {\bibfnamefont
  {S.-W.}\ \bibnamefont {Tsai}},\ and\ \bibinfo {author} {\bibfnamefont
  {Y.}~\bibnamefont {Meurice}},\ }\bibfield  {title} {\bibinfo {title}
  {Progress towards quantum simulating the classical $\text{O}(2)$ model},\
  }\href {https://doi.org/10.1103/PhysRevA.90.063603} {\bibfield  {journal}
  {\bibinfo  {journal} {Phys. Rev. A}\ }\textbf {\bibinfo {volume} {90}},\
  \bibinfo {pages} {063603} (\bibinfo {year} {2014})}\BibitemShut {NoStop}%
\bibitem [{\citenamefont {Haegeman}\ \emph {et~al.}(2015)\citenamefont
  {Haegeman}, \citenamefont {Van~Acoleyen}, \citenamefont {Schuch},
  \citenamefont {Cirac},\ and\ \citenamefont
  {Verstraete}}]{haegeman2015gauging}%
  \BibitemOpen
  \bibfield  {author} {\bibinfo {author} {\bibfnamefont {J.}~\bibnamefont
  {Haegeman}}, \bibinfo {author} {\bibfnamefont {K.}~\bibnamefont
  {Van~Acoleyen}}, \bibinfo {author} {\bibfnamefont {N.}~\bibnamefont
  {Schuch}}, \bibinfo {author} {\bibfnamefont {J.~I.}\ \bibnamefont {Cirac}},\
  and\ \bibinfo {author} {\bibfnamefont {F.}~\bibnamefont {Verstraete}},\
  }\bibfield  {title} {\bibinfo {title} {Gauging quantum states: From global to
  local symmetries in many-body systems},\ }\href
  {https://doi.org/10.1103/PhysRevX.5.011024} {\bibfield  {journal} {\bibinfo
  {journal} {Phys. Rev. X}\ }\textbf {\bibinfo {volume} {5}},\ \bibinfo {pages}
  {011024} (\bibinfo {year} {2015})}\BibitemShut {NoStop}%
\bibitem [{\citenamefont {Kuramashi}\ and\ \citenamefont
  {Yoshimura}(2019)}]{kuramashi2019three}%
  \BibitemOpen
  \bibfield  {author} {\bibinfo {author} {\bibfnamefont {Y.}~\bibnamefont
  {Kuramashi}}\ and\ \bibinfo {author} {\bibfnamefont {Y.}~\bibnamefont
  {Yoshimura}},\ }\bibfield  {title} {\bibinfo {title} {Three-dimensional
  finite temperature ${\Z}_2$ gauge theory with tensor network scheme},\ }\href
  {https://doi.org/10.1007/JHEP08(2019)023} {\bibfield  {journal} {\bibinfo
  {journal} {Journal of High Energy Physics}\ }\textbf {\bibinfo {volume}
  {2019}},\ \bibinfo {pages} {1} (\bibinfo {year} {2019})}\BibitemShut
  {NoStop}%
\bibitem [{\citenamefont {Emonts}\ and\ \citenamefont
  {Zohar}(2020)}]{emonts2020gauss}%
  \BibitemOpen
  \bibfield  {author} {\bibinfo {author} {\bibfnamefont {P.}~\bibnamefont
  {Emonts}}\ and\ \bibinfo {author} {\bibfnamefont {E.}~\bibnamefont {Zohar}},\
  }\bibfield  {title} {\bibinfo {title} {{Gauss law, minimal coupling and
  fermionic PEPS for lattice gauge theories}},\ }\href
  {https://doi.org/10.21468/SciPostPhysLectNotes.12} {\bibfield  {journal}
  {\bibinfo  {journal} {SciPost Phys. Lect. Notes}\ ,\ \bibinfo {pages} {12}}
  (\bibinfo {year} {2020})}\BibitemShut {NoStop}%
\bibitem [{\citenamefont {Felser}\ \emph {et~al.}(2020)\citenamefont {Felser},
  \citenamefont {Silvi}, \citenamefont {Collura},\ and\ \citenamefont
  {Montangero}}]{felser2020}%
  \BibitemOpen
  \bibfield  {author} {\bibinfo {author} {\bibfnamefont {T.}~\bibnamefont
  {Felser}}, \bibinfo {author} {\bibfnamefont {P.}~\bibnamefont {Silvi}},
  \bibinfo {author} {\bibfnamefont {M.}~\bibnamefont {Collura}},\ and\ \bibinfo
  {author} {\bibfnamefont {S.}~\bibnamefont {Montangero}},\ }\bibfield  {title}
  {\bibinfo {title} {Two-dimensional quantum-link lattice quantum
  electrodynamics at finite density},\ }\href
  {https://doi.org/10.1103/PhysRevX.10.041040} {\bibfield  {journal} {\bibinfo
  {journal} {Phys. Rev. X}\ }\textbf {\bibinfo {volume} {10}},\ \bibinfo
  {pages} {041040} (\bibinfo {year} {2020})}\BibitemShut {NoStop}%
\bibitem [{\citenamefont {Magnifico}\ \emph {et~al.}(2021)\citenamefont
  {Magnifico}, \citenamefont {Felser}, \citenamefont {Silvi},\ and\
  \citenamefont {Montangero}}]{magnifico2021lattice}%
  \BibitemOpen
  \bibfield  {author} {\bibinfo {author} {\bibfnamefont {G.}~\bibnamefont
  {Magnifico}}, \bibinfo {author} {\bibfnamefont {T.}~\bibnamefont {Felser}},
  \bibinfo {author} {\bibfnamefont {P.}~\bibnamefont {Silvi}},\ and\ \bibinfo
  {author} {\bibfnamefont {S.}~\bibnamefont {Montangero}},\ }\bibfield  {title}
  {\bibinfo {title} {Lattice quantum electrodynamics in (3+1)-dimensions at
  finite density with tensor networks},\ }\href
  {https://doi.org/10.1038/s41467-021-23646-3} {\bibfield  {journal} {\bibinfo
  {journal} {Nature communications}\ }\textbf {\bibinfo {volume} {12}},\
  \bibinfo {pages} {3600} (\bibinfo {year} {2021})}\BibitemShut {NoStop}%
\bibitem [{\citenamefont {Meurice}\ \emph {et~al.}(2022)\citenamefont
  {Meurice}, \citenamefont {Sakai},\ and\ \citenamefont
  {Unmuth-Yockey}}]{meurice2022}%
  \BibitemOpen
  \bibfield  {author} {\bibinfo {author} {\bibfnamefont {Y.}~\bibnamefont
  {Meurice}}, \bibinfo {author} {\bibfnamefont {R.}~\bibnamefont {Sakai}},\
  and\ \bibinfo {author} {\bibfnamefont {J.}~\bibnamefont {Unmuth-Yockey}},\
  }\bibfield  {title} {\bibinfo {title} {Tensor lattice field theory for
  renormalization and quantum computing},\ }\href
  {https://doi.org/10.1103/RevModPhys.94.025005} {\bibfield  {journal}
  {\bibinfo  {journal} {Rev. Mod. Phys.}\ }\textbf {\bibinfo {volume} {94}},\
  \bibinfo {pages} {025005} (\bibinfo {year} {2022})}\BibitemShut {NoStop}%
\bibitem [{\citenamefont {Cataldi}\ \emph {et~al.}(2024)\citenamefont
  {Cataldi}, \citenamefont {Magnifico}, \citenamefont {Silvi},\ and\
  \citenamefont {Montangero}}]{cataldi2024}%
  \BibitemOpen
  \bibfield  {author} {\bibinfo {author} {\bibfnamefont {G.}~\bibnamefont
  {Cataldi}}, \bibinfo {author} {\bibfnamefont {G.}~\bibnamefont {Magnifico}},
  \bibinfo {author} {\bibfnamefont {P.}~\bibnamefont {Silvi}},\ and\ \bibinfo
  {author} {\bibfnamefont {S.}~\bibnamefont {Montangero}},\ }\bibfield  {title}
  {\bibinfo {title} {Simulating $(2+1)\mathrm{D}$ \text{SU(2) Yang-Mills}
  lattice gauge theory at finite density with tensor networks},\ }\href
  {https://doi.org/10.1103/PhysRevResearch.6.033057} {\bibfield  {journal}
  {\bibinfo  {journal} {Phys. Rev. Res.}\ }\textbf {\bibinfo {volume} {6}},\
  \bibinfo {pages} {033057} (\bibinfo {year} {2024})}\BibitemShut {NoStop}%
\bibitem [{\citenamefont {Byrnes}\ \emph
  {et~al.}(2002{\natexlab{a}})\citenamefont {Byrnes}, \citenamefont
  {Sriganesh}, \citenamefont {Bursill},\ and\ \citenamefont
  {Hamer}}]{LGT_DMRG1}%
  \BibitemOpen
  \bibfield  {author} {\bibinfo {author} {\bibfnamefont {T.}~\bibnamefont
  {Byrnes}}, \bibinfo {author} {\bibfnamefont {P.}~\bibnamefont {Sriganesh}},
  \bibinfo {author} {\bibfnamefont {R.}~\bibnamefont {Bursill}},\ and\ \bibinfo
  {author} {\bibfnamefont {C.}~\bibnamefont {Hamer}},\ }\bibfield  {title}
  {\bibinfo {title} {Density matrix renormalisation group approach to the
  massive schwinger model},\ }\href
  {https://doi.org/https://doi.org/10.1016/S0920-5632(02)01416-0} {\bibfield
  {journal} {\bibinfo  {journal} {Nuclear Physics B - Proceedings Supplements}\
  }\textbf {\bibinfo {volume} {109}},\ \bibinfo {pages} {202} (\bibinfo {year}
  {2002}{\natexlab{a}})}\BibitemShut {NoStop}%
\bibitem [{\citenamefont {Byrnes}\ \emph
  {et~al.}(2002{\natexlab{b}})\citenamefont {Byrnes}, \citenamefont
  {Sriganesh}, \citenamefont {Bursill},\ and\ \citenamefont
  {Hamer}}]{LGT_DMRG2}%
  \BibitemOpen
  \bibfield  {author} {\bibinfo {author} {\bibfnamefont {T.~M.~R.}\
  \bibnamefont {Byrnes}}, \bibinfo {author} {\bibfnamefont {P.}~\bibnamefont
  {Sriganesh}}, \bibinfo {author} {\bibfnamefont {R.~J.}\ \bibnamefont
  {Bursill}},\ and\ \bibinfo {author} {\bibfnamefont {C.~J.}\ \bibnamefont
  {Hamer}},\ }\bibfield  {title} {\bibinfo {title} {Density matrix
  renormalization group approach to the massive schwinger model},\ }\href
  {https://doi.org/10.1103/PhysRevD.66.013002} {\bibfield  {journal} {\bibinfo
  {journal} {Phys. Rev. D}\ }\textbf {\bibinfo {volume} {66}},\ \bibinfo
  {pages} {013002} (\bibinfo {year} {2002}{\natexlab{b}})}\BibitemShut
  {NoStop}%
\bibitem [{\citenamefont {Ba{\~n}uls}\ \emph {et~al.}(2013)\citenamefont
  {Ba{\~n}uls}, \citenamefont {Cichy}, \citenamefont {Cirac},\ and\
  \citenamefont {Jansen}}]{bamuls2013the}%
  \BibitemOpen
  \bibfield  {author} {\bibinfo {author} {\bibfnamefont {M.~C.}\ \bibnamefont
  {Ba{\~n}uls}}, \bibinfo {author} {\bibfnamefont {K.}~\bibnamefont {Cichy}},
  \bibinfo {author} {\bibfnamefont {J.~I.}\ \bibnamefont {Cirac}},\ and\
  \bibinfo {author} {\bibfnamefont {K.}~\bibnamefont {Jansen}},\ }\bibfield
  {title} {\bibinfo {title} {The mass spectrum of the schwinger model with
  matrix product states},\ }\href {https://doi.org/10.1007/JHEP11(2013)158}
  {\bibfield  {journal} {\bibinfo  {journal} {Journal of High Energy Physics}\
  }\textbf {\bibinfo {volume} {2013}},\ \bibinfo {pages} {1} (\bibinfo {year}
  {2013})}\BibitemShut {NoStop}%
\bibitem [{\citenamefont {Rico}\ \emph {et~al.}(2014)\citenamefont {Rico},
  \citenamefont {Pichler}, \citenamefont {Dalmonte}, \citenamefont {Zoller},\
  and\ \citenamefont {Montangero}}]{rico2014tensor}%
  \BibitemOpen
  \bibfield  {author} {\bibinfo {author} {\bibfnamefont {E.}~\bibnamefont
  {Rico}}, \bibinfo {author} {\bibfnamefont {T.}~\bibnamefont {Pichler}},
  \bibinfo {author} {\bibfnamefont {M.}~\bibnamefont {Dalmonte}}, \bibinfo
  {author} {\bibfnamefont {P.}~\bibnamefont {Zoller}},\ and\ \bibinfo {author}
  {\bibfnamefont {S.}~\bibnamefont {Montangero}},\ }\bibfield  {title}
  {\bibinfo {title} {Tensor networks for lattice gauge theories and atomic
  quantum simulation},\ }\href {https://doi.org/10.1103/PhysRevLett.112.201601}
  {\bibfield  {journal} {\bibinfo  {journal} {Phys. Rev. Lett.}\ }\textbf
  {\bibinfo {volume} {112}},\ \bibinfo {pages} {201601} (\bibinfo {year}
  {2014})}\BibitemShut {NoStop}%
\bibitem [{\citenamefont {Carmen~Bañuls}\ and\ \citenamefont
  {Cichy}(2020)}]{banus2020review}%
  \BibitemOpen
  \bibfield  {author} {\bibinfo {author} {\bibfnamefont {M.}~\bibnamefont
  {Carmen~Bañuls}}\ and\ \bibinfo {author} {\bibfnamefont {K.}~\bibnamefont
  {Cichy}},\ }\bibfield  {title} {\bibinfo {title} {Review on novel methods for
  lattice gauge theories},\ }\href {https://doi.org/10.1088/1361-6633/ab6311}
  {\bibfield  {journal} {\bibinfo  {journal} {Reports on Progress in Physics}\
  }\textbf {\bibinfo {volume} {83}},\ \bibinfo {pages} {024401} (\bibinfo
  {year} {2020})}\BibitemShut {NoStop}%
\bibitem [{\citenamefont {Verstraete}\ and\ \citenamefont
  {Cirac}(2004)}]{PEPS2004}%
  \BibitemOpen
  \bibfield  {author} {\bibinfo {author} {\bibfnamefont {F.}~\bibnamefont
  {Verstraete}}\ and\ \bibinfo {author} {\bibfnamefont {J.~I.}\ \bibnamefont
  {Cirac}},\ }\bibfield  {title} {\bibinfo {title} {Renormalization algorithms
  for quantum-many body systems in two and higher dimensions},\ }\href
  {https://doi.org/10.48550/arXiv.cond-mat/0407066} {\bibfield  {journal}
  {\bibinfo  {journal} {arXiv:cond-mat/0407066}\ } (\bibinfo {year}
  {2004})}\BibitemShut {NoStop}%
\bibitem [{\citenamefont {Zohar}\ \emph {et~al.}(2015)\citenamefont {Zohar},
  \citenamefont {Burrello}, \citenamefont {Wahl},\ and\ \citenamefont
  {Cirac}}]{zohar2015fermion}%
  \BibitemOpen
  \bibfield  {author} {\bibinfo {author} {\bibfnamefont {E.}~\bibnamefont
  {Zohar}}, \bibinfo {author} {\bibfnamefont {M.}~\bibnamefont {Burrello}},
  \bibinfo {author} {\bibfnamefont {T.~B.}\ \bibnamefont {Wahl}},\ and\
  \bibinfo {author} {\bibfnamefont {J.~I.}\ \bibnamefont {Cirac}},\ }\bibfield
  {title} {\bibinfo {title} {Fermionic projected entangled pair states and
  local \text{U(1)} gauge theories},\ }\href
  {https://doi.org/https://doi.org/10.1016/j.aop.2015.10.009} {\bibfield
  {journal} {\bibinfo  {journal} {Annals of Physics}\ }\textbf {\bibinfo
  {volume} {363}},\ \bibinfo {pages} {385} (\bibinfo {year}
  {2015})}\BibitemShut {NoStop}%
\bibitem [{\citenamefont {Zohar}\ and\ \citenamefont
  {Burrello}(2016)}]{zohar2016building}%
  \BibitemOpen
  \bibfield  {author} {\bibinfo {author} {\bibfnamefont {E.}~\bibnamefont
  {Zohar}}\ and\ \bibinfo {author} {\bibfnamefont {M.}~\bibnamefont
  {Burrello}},\ }\bibfield  {title} {\bibinfo {title} {Building projected
  entangled pair states with a local gauge symmetry},\ }\href
  {https://doi.org/10.1088/1367-2630/18/4/043008} {\bibfield  {journal}
  {\bibinfo  {journal} {New Journal of Physics}\ }\textbf {\bibinfo {volume}
  {18}},\ \bibinfo {pages} {043008} (\bibinfo {year} {2016})}\BibitemShut
  {NoStop}%
\bibitem [{\citenamefont {Blanik}\ \emph {et~al.}(2024)\citenamefont {Blanik},
  \citenamefont {Garre-Rubio}, \citenamefont {Moln\'ar},\ and\ \citenamefont
  {Zohar}}]{Blanik2024}%
  \BibitemOpen
  \bibfield  {author} {\bibinfo {author} {\bibfnamefont {D.}~\bibnamefont
  {Blanik}}, \bibinfo {author} {\bibfnamefont {J.}~\bibnamefont {Garre-Rubio}},
  \bibinfo {author} {\bibfnamefont {A.}~\bibnamefont {Moln\'ar}},\ and\
  \bibinfo {author} {\bibfnamefont {E.}~\bibnamefont {Zohar}},\ }\bibfield
  {title} {\bibinfo {title} {Internal structure of gauge-invariant projected
  entangled pair states},\ }\href {https://doi.org/10.48550/arXiv.2410.18947}
  {\bibfield  {journal} {\bibinfo  {journal} {arXiv:2410.18947}\ } (\bibinfo
  {year} {2024})}\BibitemShut {NoStop}%
\bibitem [{\citenamefont {Roose}\ and\ \citenamefont
  {Zohar}(2024)}]{Roose2024}%
  \BibitemOpen
  \bibfield  {author} {\bibinfo {author} {\bibfnamefont {G.}~\bibnamefont
  {Roose}}\ and\ \bibinfo {author} {\bibfnamefont {E.}~\bibnamefont {Zohar}},\
  }\bibfield  {title} {\bibinfo {title} {Superposing and gauging fermionic
  gaussian projected entangled pair states to get exact lattice gauge theory
  groundstates},\ }\href {https://doi.org/10.48550/arXiv.2412.01737} {\bibfield
   {journal} {\bibinfo  {journal} {arXiv:2412.01737}\ } (\bibinfo {year}
  {2024})}\BibitemShut {NoStop}%
\bibitem [{\citenamefont {Zohar}\ and\ \citenamefont
  {Cirac}(2018)}]{zohar2018combining}%
  \BibitemOpen
  \bibfield  {author} {\bibinfo {author} {\bibfnamefont {E.}~\bibnamefont
  {Zohar}}\ and\ \bibinfo {author} {\bibfnamefont {J.~I.}\ \bibnamefont
  {Cirac}},\ }\bibfield  {title} {\bibinfo {title} {Combining tensor networks
  with monte carlo methods for lattice gauge theories},\ }\href
  {https://doi.org/10.1103/PhysRevD.97.034510} {\bibfield  {journal} {\bibinfo
  {journal} {Phys. Rev. D}\ }\textbf {\bibinfo {volume} {97}},\ \bibinfo
  {pages} {034510} (\bibinfo {year} {2018})}\BibitemShut {NoStop}%
\bibitem [{\citenamefont {Emonts}\ \emph {et~al.}(2023)\citenamefont {Emonts},
  \citenamefont {Kelman}, \citenamefont {Borla}, \citenamefont {Moroz},
  \citenamefont {Gazit},\ and\ \citenamefont {Zohar}}]{emonts2023finding}%
  \BibitemOpen
  \bibfield  {author} {\bibinfo {author} {\bibfnamefont {P.}~\bibnamefont
  {Emonts}}, \bibinfo {author} {\bibfnamefont {A.}~\bibnamefont {Kelman}},
  \bibinfo {author} {\bibfnamefont {U.}~\bibnamefont {Borla}}, \bibinfo
  {author} {\bibfnamefont {S.}~\bibnamefont {Moroz}}, \bibinfo {author}
  {\bibfnamefont {S.}~\bibnamefont {Gazit}},\ and\ \bibinfo {author}
  {\bibfnamefont {E.}~\bibnamefont {Zohar}},\ }\bibfield  {title} {\bibinfo
  {title} {Finding the ground state of a lattice gauge theory with fermionic
  tensor networks: A $2+1 d$ ${\Z}_{2}$ demonstration},\ }\href
  {https://doi.org/10.1103/PhysRevD.107.014505} {\bibfield  {journal} {\bibinfo
   {journal} {Phys. Rev. D}\ }\textbf {\bibinfo {volume} {107}},\ \bibinfo
  {pages} {014505} (\bibinfo {year} {2023})}\BibitemShut {NoStop}%
\bibitem [{\citenamefont {Emonts}\ \emph {et~al.}(2020)\citenamefont {Emonts},
  \citenamefont {Ba\~nuls}, \citenamefont {Cirac},\ and\ \citenamefont
  {Zohar}}]{emonts2020varational}%
  \BibitemOpen
  \bibfield  {author} {\bibinfo {author} {\bibfnamefont {P.}~\bibnamefont
  {Emonts}}, \bibinfo {author} {\bibfnamefont {M.~C.}\ \bibnamefont
  {Ba\~nuls}}, \bibinfo {author} {\bibfnamefont {I.}~\bibnamefont {Cirac}},\
  and\ \bibinfo {author} {\bibfnamefont {E.}~\bibnamefont {Zohar}},\ }\bibfield
   {title} {\bibinfo {title} {Variational \text{Monte Carlo} simulation with
  tensor networks of a pure ${Z}_{3}$ gauge theory in \text{2+1 D}},\ }\href
  {https://doi.org/10.1103/PhysRevD.102.074501} {\bibfield  {journal} {\bibinfo
   {journal} {Phys. Rev. D}\ }\textbf {\bibinfo {volume} {102}},\ \bibinfo
  {pages} {074501} (\bibinfo {year} {2020})}\BibitemShut {NoStop}%
\bibitem [{\citenamefont {Kelman}\ \emph {et~al.}(2024)\citenamefont {Kelman},
  \citenamefont {Borla}, \citenamefont {Gomelski}, \citenamefont {Elyovich},
  \citenamefont {Roose}, \citenamefont {Emonts},\ and\ \citenamefont
  {Zohar}}]{kelman2024gauged}%
  \BibitemOpen
  \bibfield  {author} {\bibinfo {author} {\bibfnamefont {A.}~\bibnamefont
  {Kelman}}, \bibinfo {author} {\bibfnamefont {U.}~\bibnamefont {Borla}},
  \bibinfo {author} {\bibfnamefont {I.}~\bibnamefont {Gomelski}}, \bibinfo
  {author} {\bibfnamefont {J.}~\bibnamefont {Elyovich}}, \bibinfo {author}
  {\bibfnamefont {G.}~\bibnamefont {Roose}}, \bibinfo {author} {\bibfnamefont
  {P.}~\bibnamefont {Emonts}},\ and\ \bibinfo {author} {\bibfnamefont
  {E.}~\bibnamefont {Zohar}},\ }\bibfield  {title} {\bibinfo {title} {Gauged
  gaussian projected entangled pair states: A high dimensional tensor network
  formulation for lattice gauge theories},\ }\href
  {https://doi.org/10.1103/PhysRevD.110.054511} {\bibfield  {journal} {\bibinfo
   {journal} {Phys. Rev. D}\ }\textbf {\bibinfo {volume} {110}},\ \bibinfo
  {pages} {054511} (\bibinfo {year} {2024})}\BibitemShut {NoStop}%
\bibitem [{\citenamefont {Robaina}\ \emph {et~al.}(2021)\citenamefont
  {Robaina}, \citenamefont {Ba\~nuls},\ and\ \citenamefont
  {Cirac}}]{iPEPS_LGT2021}%
  \BibitemOpen
  \bibfield  {author} {\bibinfo {author} {\bibfnamefont {D.}~\bibnamefont
  {Robaina}}, \bibinfo {author} {\bibfnamefont {M.~C.}\ \bibnamefont
  {Ba\~nuls}},\ and\ \bibinfo {author} {\bibfnamefont {J.~I.}\ \bibnamefont
  {Cirac}},\ }\bibfield  {title} {\bibinfo {title} {Simulating \text{2+1 D}
  ${\Z}_{3}$ lattice gauge theory with an infinite projected entangled-pair
  state},\ }\href {https://doi.org/10.1103/PhysRevLett.126.050401} {\bibfield
  {journal} {\bibinfo  {journal} {Phys. Rev. Lett.}\ }\textbf {\bibinfo
  {volume} {126}},\ \bibinfo {pages} {050401} (\bibinfo {year}
  {2021})}\BibitemShut {NoStop}%
\bibitem [{\citenamefont {Kogut}\ and\ \citenamefont
  {Susskind}(1975)}]{kogut1975}%
  \BibitemOpen
  \bibfield  {author} {\bibinfo {author} {\bibfnamefont {J.}~\bibnamefont
  {Kogut}}\ and\ \bibinfo {author} {\bibfnamefont {L.}~\bibnamefont
  {Susskind}},\ }\bibfield  {title} {\bibinfo {title} {Hamiltonian formulation
  of wilson's lattice gauge theories},\ }\href
  {https://doi.org/10.1103/PhysRevD.11.395} {\bibfield  {journal} {\bibinfo
  {journal} {Phys. Rev. D}\ }\textbf {\bibinfo {volume} {11}},\ \bibinfo
  {pages} {395} (\bibinfo {year} {1975})}\BibitemShut {NoStop}%
\bibitem [{\citenamefont {Sandvik}\ and\ \citenamefont
  {Vidal}(2007)}]{sandvik2007}%
  \BibitemOpen
  \bibfield  {author} {\bibinfo {author} {\bibfnamefont {A.~W.}\ \bibnamefont
  {Sandvik}}\ and\ \bibinfo {author} {\bibfnamefont {G.}~\bibnamefont
  {Vidal}},\ }\bibfield  {title} {\bibinfo {title} {Variational quantum
  \text{Monte} \text{Carlo} simulations with tensor-network states},\ }\href
  {https://doi.org/10.1103/PhysRevLett.99.220602} {\bibfield  {journal}
  {\bibinfo  {journal} {Phys. Rev. Lett.}\ }\textbf {\bibinfo {volume} {99}},\
  \bibinfo {pages} {220602} (\bibinfo {year} {2007})}\BibitemShut {NoStop}%
\bibitem [{\citenamefont {Schuch}\ \emph {et~al.}(2008)\citenamefont {Schuch},
  \citenamefont {Wolf}, \citenamefont {Verstraete},\ and\ \citenamefont
  {Cirac}}]{schuch2008}%
  \BibitemOpen
  \bibfield  {author} {\bibinfo {author} {\bibfnamefont {N.}~\bibnamefont
  {Schuch}}, \bibinfo {author} {\bibfnamefont {M.~M.}\ \bibnamefont {Wolf}},
  \bibinfo {author} {\bibfnamefont {F.}~\bibnamefont {Verstraete}},\ and\
  \bibinfo {author} {\bibfnamefont {J.~I.}\ \bibnamefont {Cirac}},\ }\bibfield
  {title} {\bibinfo {title} {Simulation of quantum many-body systems with
  strings of operators and \text{Monte} \text{Carlo} tensor contractions},\
  }\href {http://dx.doi.org/10.1103/PhysRevLett.100.040501} {\bibfield
  {journal} {\bibinfo  {journal} {Phys. Rev. Lett.}\ }\textbf {\bibinfo
  {volume} {100}},\ \bibinfo {pages} {040501} (\bibinfo {year}
  {2008})}\BibitemShut {NoStop}%
\bibitem [{\citenamefont {Wang}\ \emph {et~al.}(2011)\citenamefont {Wang},
  \citenamefont {Pi\ifmmode~\check{z}\else \v{z}\fi{}orn},\ and\ \citenamefont
  {Verstraete}}]{wang2011}%
  \BibitemOpen
  \bibfield  {author} {\bibinfo {author} {\bibfnamefont {L.}~\bibnamefont
  {Wang}}, \bibinfo {author} {\bibfnamefont {I.}~\bibnamefont
  {Pi\ifmmode~\check{z}\else \v{z}\fi{}orn}},\ and\ \bibinfo {author}
  {\bibfnamefont {F.}~\bibnamefont {Verstraete}},\ }\bibfield  {title}
  {\bibinfo {title} {\text{Monte} \text{Carlo} simulation with tensor network
  states},\ }\href {https://doi.org/10.1103/PhysRevB.83.134421} {\bibfield
  {journal} {\bibinfo  {journal} {Phys. Rev. B}\ }\textbf {\bibinfo {volume}
  {83}},\ \bibinfo {pages} {134421} (\bibinfo {year} {2011})}\BibitemShut
  {NoStop}%
\bibitem [{\citenamefont {Liu}\ \emph {et~al.}(2017)\citenamefont {Liu},
  \citenamefont {Dong}, \citenamefont {Han}, \citenamefont {Guo},\ and\
  \citenamefont {He}}]{liu2017}%
  \BibitemOpen
  \bibfield  {author} {\bibinfo {author} {\bibfnamefont {W.-Y.}\ \bibnamefont
  {Liu}}, \bibinfo {author} {\bibfnamefont {S.-J.}\ \bibnamefont {Dong}},
  \bibinfo {author} {\bibfnamefont {Y.-J.}\ \bibnamefont {Han}}, \bibinfo
  {author} {\bibfnamefont {G.-C.}\ \bibnamefont {Guo}},\ and\ \bibinfo {author}
  {\bibfnamefont {L.}~\bibnamefont {He}},\ }\bibfield  {title} {\bibinfo
  {title} {Gradient optimization of finite projected entangled pair states},\
  }\href {https://doi.org/10.1103/PhysRevB.95.195154} {\bibfield  {journal}
  {\bibinfo  {journal} {Phys. Rev. B}\ }\textbf {\bibinfo {volume} {95}},\
  \bibinfo {pages} {195154} (\bibinfo {year} {2017})}\BibitemShut {NoStop}%
\bibitem [{\citenamefont {Liu}\ \emph {et~al.}(2021)\citenamefont {Liu},
  \citenamefont {Huang}, \citenamefont {Gong},\ and\ \citenamefont
  {Gu}}]{liu2021}%
  \BibitemOpen
  \bibfield  {author} {\bibinfo {author} {\bibfnamefont {W.-Y.}\ \bibnamefont
  {Liu}}, \bibinfo {author} {\bibfnamefont {Y.-Z.}\ \bibnamefont {Huang}},
  \bibinfo {author} {\bibfnamefont {S.-S.}\ \bibnamefont {Gong}},\ and\
  \bibinfo {author} {\bibfnamefont {Z.-C.}\ \bibnamefont {Gu}},\ }\bibfield
  {title} {\bibinfo {title} {Accurate simulation for finite projected entangled
  pair states in two dimensions},\ }\href
  {https://doi.org/10.1103/PhysRevB.103.235155} {\bibfield  {journal} {\bibinfo
   {journal} {Phys. Rev. B}\ }\textbf {\bibinfo {volume} {103}},\ \bibinfo
  {pages} {235155} (\bibinfo {year} {2021})}\BibitemShut {NoStop}%
\bibitem [{\citenamefont {Liu}\ \emph {et~al.}(2024{\natexlab{a}})\citenamefont
  {Liu}, \citenamefont {Du}, \citenamefont {Peng}, \citenamefont {Gray},\ and\
  \citenamefont {Chan}}]{liu2024tnf}%
  \BibitemOpen
  \bibfield  {author} {\bibinfo {author} {\bibfnamefont {W.-Y.}\ \bibnamefont
  {Liu}}, \bibinfo {author} {\bibfnamefont {S.-J.}\ \bibnamefont {Du}},
  \bibinfo {author} {\bibfnamefont {R.}~\bibnamefont {Peng}}, \bibinfo {author}
  {\bibfnamefont {J.}~\bibnamefont {Gray}},\ and\ \bibinfo {author}
  {\bibfnamefont {G.~K.-L.}\ \bibnamefont {Chan}},\ }\bibfield  {title}
  {\bibinfo {title} {Tensor network computations that capture strict
  variationality, volume law behavior, and the efficient representation of
  neural network states},\ }\href
  {https://doi.org/10.1103/PhysRevLett.133.260404} {\bibfield  {journal}
  {\bibinfo  {journal} {Phys. Rev. Lett.}\ }\textbf {\bibinfo {volume} {133}},\
  \bibinfo {pages} {260404} (\bibinfo {year} {2024}{\natexlab{a}})}\BibitemShut
  {NoStop}%
\bibitem [{\citenamefont {Liu}\ \emph {et~al.}(2025)\citenamefont {Liu},
  \citenamefont {Zhai}, \citenamefont {Peng}, \citenamefont {Gu},\ and\
  \citenamefont {Chan}}]{liu2025}%
  \BibitemOpen
  \bibfield  {author} {\bibinfo {author} {\bibfnamefont {W.-Y.}\ \bibnamefont
  {Liu}}, \bibinfo {author} {\bibfnamefont {H.}~\bibnamefont {Zhai}}, \bibinfo
  {author} {\bibfnamefont {R.}~\bibnamefont {Peng}}, \bibinfo {author}
  {\bibfnamefont {Z.-C.}\ \bibnamefont {Gu}},\ and\ \bibinfo {author}
  {\bibfnamefont {G.~K.-L.}\ \bibnamefont {Chan}},\ }\bibfield  {title}
  {\bibinfo {title} {Accurate simulation of the \text{Hubbard} model with
  finite fermionic projected entangled pair states},\ }\href
  {https://doi.org/10.48550/arXiv.2502.13454} {\bibfield  {journal} {\bibinfo
  {journal} {arXiv:2502.13454}\ } (\bibinfo {year} {2025})}\BibitemShut
  {NoStop}%
\bibitem [{\citenamefont {Liu}\ \emph {et~al.}(2018)\citenamefont {Liu},
  \citenamefont {Dong}, \citenamefont {Wang}, \citenamefont {Han},
  \citenamefont {An}, \citenamefont {Guo},\ and\ \citenamefont
  {He}}]{liu2018gapless}%
  \BibitemOpen
  \bibfield  {author} {\bibinfo {author} {\bibfnamefont {W.-Y.}\ \bibnamefont
  {Liu}}, \bibinfo {author} {\bibfnamefont {S.}~\bibnamefont {Dong}}, \bibinfo
  {author} {\bibfnamefont {C.}~\bibnamefont {Wang}}, \bibinfo {author}
  {\bibfnamefont {Y.}~\bibnamefont {Han}}, \bibinfo {author} {\bibfnamefont
  {H.}~\bibnamefont {An}}, \bibinfo {author} {\bibfnamefont {G.-C.}\
  \bibnamefont {Guo}},\ and\ \bibinfo {author} {\bibfnamefont {L.}~\bibnamefont
  {He}},\ }\bibfield  {title} {\bibinfo {title} {Gapless spin liquid ground
  state of the spin-$\frac{1}{2}$ ${J}_{1}\ensuremath{-}{J}_{2}$ heisenberg
  model on square lattices},\ }\href
  {https://doi.org/10.1103/PhysRevB.98.241109} {\bibfield  {journal} {\bibinfo
  {journal} {Phys. Rev. B}\ }\textbf {\bibinfo {volume} {98}},\ \bibinfo
  {pages} {241109} (\bibinfo {year} {2018})}\BibitemShut {NoStop}%
\bibitem [{\citenamefont {Liu}\ \emph {et~al.}(2022{\natexlab{a}})\citenamefont
  {Liu}, \citenamefont {Gong}, \citenamefont {Li}, \citenamefont {Poilblanc},
  \citenamefont {Chen},\ and\ \citenamefont {Gu}}]{liu2022gapless}%
  \BibitemOpen
  \bibfield  {author} {\bibinfo {author} {\bibfnamefont {W.-Y.}\ \bibnamefont
  {Liu}}, \bibinfo {author} {\bibfnamefont {S.-S.}\ \bibnamefont {Gong}},
  \bibinfo {author} {\bibfnamefont {Y.-B.}\ \bibnamefont {Li}}, \bibinfo
  {author} {\bibfnamefont {D.}~\bibnamefont {Poilblanc}}, \bibinfo {author}
  {\bibfnamefont {W.-Q.}\ \bibnamefont {Chen}},\ and\ \bibinfo {author}
  {\bibfnamefont {Z.-C.}\ \bibnamefont {Gu}},\ }\bibfield  {title} {\bibinfo
  {title} {Gapless quantum spin liquid and global phase diagram of the spin-1/2
  ${J}_{1}-{J}_{2}$ square antiferromagnetic \text{Heisenberg} model},\ }\href
  {https://doi.org/https://doi.org/10.1016/j.scib.2022.03.010} {\bibfield
  {journal} {\bibinfo  {journal} {Science Bulletin}\ }\textbf {\bibinfo
  {volume} {67}},\ \bibinfo {pages} {1034} (\bibinfo {year}
  {2022}{\natexlab{a}})}\BibitemShut {NoStop}%
\bibitem [{\citenamefont {Liu}\ \emph {et~al.}(2022{\natexlab{b}})\citenamefont
  {Liu}, \citenamefont {Hasik}, \citenamefont {Gong}, \citenamefont
  {Poilblanc}, \citenamefont {Chen},\ and\ \citenamefont
  {Gu}}]{liu2022emergence}%
  \BibitemOpen
  \bibfield  {author} {\bibinfo {author} {\bibfnamefont {W.-Y.}\ \bibnamefont
  {Liu}}, \bibinfo {author} {\bibfnamefont {J.}~\bibnamefont {Hasik}}, \bibinfo
  {author} {\bibfnamefont {S.-S.}\ \bibnamefont {Gong}}, \bibinfo {author}
  {\bibfnamefont {D.}~\bibnamefont {Poilblanc}}, \bibinfo {author}
  {\bibfnamefont {W.-Q.}\ \bibnamefont {Chen}},\ and\ \bibinfo {author}
  {\bibfnamefont {Z.-C.}\ \bibnamefont {Gu}},\ }\bibfield  {title} {\bibinfo
  {title} {Emergence of gapless quantum spin liquid from deconfined quantum
  critical point},\ }\href {https://doi.org/10.1103/PhysRevX.12.031039}
  {\bibfield  {journal} {\bibinfo  {journal} {Phys. Rev. X}\ }\textbf {\bibinfo
  {volume} {12}},\ \bibinfo {pages} {031039} (\bibinfo {year}
  {2022}{\natexlab{b}})}\BibitemShut {NoStop}%
\bibitem [{\citenamefont {Liu}\ \emph {et~al.}(2024{\natexlab{b}})\citenamefont
  {Liu}, \citenamefont {Gong}, \citenamefont {Chen},\ and\ \citenamefont
  {Gu}}]{liu2024emergent}%
  \BibitemOpen
  \bibfield  {author} {\bibinfo {author} {\bibfnamefont {W.-Y.}\ \bibnamefont
  {Liu}}, \bibinfo {author} {\bibfnamefont {S.-S.}\ \bibnamefont {Gong}},
  \bibinfo {author} {\bibfnamefont {W.-Q.}\ \bibnamefont {Chen}},\ and\
  \bibinfo {author} {\bibfnamefont {Z.-C.}\ \bibnamefont {Gu}},\ }\bibfield
  {title} {\bibinfo {title} {Emergent symmetry in quantum phase transition:
  From deconfined quantum critical point to gapless quantum spin liquid},\
  }\href {https://doi.org/https://doi.org/10.1016/j.scib.2023.11.057}
  {\bibfield  {journal} {\bibinfo  {journal} {Science Bulletin}\ }\textbf
  {\bibinfo {volume} {69}},\ \bibinfo {pages} {190} (\bibinfo {year}
  {2024}{\natexlab{b}})}\BibitemShut {NoStop}%
\bibitem [{\citenamefont {Liu}\ \emph {et~al.}(2024{\natexlab{c}})\citenamefont
  {Liu}, \citenamefont {Poilblanc}, \citenamefont {Gong}, \citenamefont
  {Chen},\ and\ \citenamefont {Gu}}]{liu2024j1j2j3}%
  \BibitemOpen
  \bibfield  {author} {\bibinfo {author} {\bibfnamefont {W.-Y.}\ \bibnamefont
  {Liu}}, \bibinfo {author} {\bibfnamefont {D.}~\bibnamefont {Poilblanc}},
  \bibinfo {author} {\bibfnamefont {S.-S.}\ \bibnamefont {Gong}}, \bibinfo
  {author} {\bibfnamefont {W.-Q.}\ \bibnamefont {Chen}},\ and\ \bibinfo
  {author} {\bibfnamefont {Z.-C.}\ \bibnamefont {Gu}},\ }\bibfield  {title}
  {\bibinfo {title} {Tensor network study of the spin-$\frac{1}{2}$
  square-lattice
  ${J}_{1}\text{\ensuremath{-}}{J}_{2}\text{\ensuremath{-}}{J}_{3}$ model:
  Incommensurate spiral order, mixed valence-bond solids, and multicritical
  points},\ }\href {https://doi.org/10.1103/PhysRevB.109.235116} {\bibfield
  {journal} {\bibinfo  {journal} {Phys. Rev. B}\ }\textbf {\bibinfo {volume}
  {109}},\ \bibinfo {pages} {235116} (\bibinfo {year}
  {2024}{\natexlab{c}})}\BibitemShut {NoStop}%
\bibitem [{\citenamefont {Liu}\ \emph {et~al.}(2024{\natexlab{d}})\citenamefont
  {Liu}, \citenamefont {Zhang}, \citenamefont {Wang}, \citenamefont {Gong},
  \citenamefont {Chen},\ and\ \citenamefont {Gu}}]{liu2024quantum}%
  \BibitemOpen
  \bibfield  {author} {\bibinfo {author} {\bibfnamefont {W.-Y.}\ \bibnamefont
  {Liu}}, \bibinfo {author} {\bibfnamefont {X.-T.}\ \bibnamefont {Zhang}},
  \bibinfo {author} {\bibfnamefont {Z.}~\bibnamefont {Wang}}, \bibinfo {author}
  {\bibfnamefont {S.-S.}\ \bibnamefont {Gong}}, \bibinfo {author}
  {\bibfnamefont {W.-Q.}\ \bibnamefont {Chen}},\ and\ \bibinfo {author}
  {\bibfnamefont {Z.-C.}\ \bibnamefont {Gu}},\ }\bibfield  {title} {\bibinfo
  {title} {Quantum criticality with emergent symmetry in the extended
  shastry-sutherland model},\ }\href
  {https://doi.org/10.1103/PhysRevLett.133.026502} {\bibfield  {journal}
  {\bibinfo  {journal} {Phys. Rev. Lett.}\ }\textbf {\bibinfo {volume} {133}},\
  \bibinfo {pages} {026502} (\bibinfo {year} {2024}{\natexlab{d}})}\BibitemShut
  {NoStop}%
\bibitem [{\citenamefont {Wu}\ and\ \citenamefont {Liu}(2025)}]{SM}%
  \BibitemOpen
  \bibfield  {author} {\bibinfo {author} {\bibfnamefont {Y.}~\bibnamefont
  {Wu}}\ and\ \bibinfo {author} {\bibfnamefont {W.-Y.}\ \bibnamefont {Liu}},\
  }\href@noop {} {\bibinfo {title} {Supplemental material}} (\bibinfo {year}
  {2025})\BibitemShut {NoStop}%
\bibitem [{\citenamefont {Sorella}(1998)}]{SR1998}%
  \BibitemOpen
  \bibfield  {author} {\bibinfo {author} {\bibfnamefont {S.}~\bibnamefont
  {Sorella}},\ }\bibfield  {title} {\bibinfo {title} {Green function monte
  carlo with stochastic reconfiguration},\ }\href
  {https://doi.org/10.1103/PhysRevLett.80.4558} {\bibfield  {journal} {\bibinfo
   {journal} {Phys. Rev. Lett.}\ }\textbf {\bibinfo {volume} {80}},\ \bibinfo
  {pages} {4558} (\bibinfo {year} {1998})}\BibitemShut {NoStop}%
\bibitem [{\citenamefont {Sorella}(2001)}]{tvmc}%
  \BibitemOpen
  \bibfield  {author} {\bibinfo {author} {\bibfnamefont {S.}~\bibnamefont
  {Sorella}},\ }\bibfield  {title} {\bibinfo {title} {Generalized lanczos
  algorithm for variational quantum monte carlo},\ }\href
  {https://doi.org/10.1103/PhysRevB.64.024512} {\bibfield  {journal} {\bibinfo
  {journal} {Phys. Rev. B}\ }\textbf {\bibinfo {volume} {64}},\ \bibinfo
  {pages} {024512} (\bibinfo {year} {2001})}\BibitemShut {NoStop}%
\bibitem [{\citenamefont {Vieijra}\ \emph {et~al.}(2021)\citenamefont
  {Vieijra}, \citenamefont {Haegeman}, \citenamefont {Verstraete},\ and\
  \citenamefont {Vanderstraeten}}]{directSam2021}%
  \BibitemOpen
  \bibfield  {author} {\bibinfo {author} {\bibfnamefont {T.}~\bibnamefont
  {Vieijra}}, \bibinfo {author} {\bibfnamefont {J.}~\bibnamefont {Haegeman}},
  \bibinfo {author} {\bibfnamefont {F.}~\bibnamefont {Verstraete}},\ and\
  \bibinfo {author} {\bibfnamefont {L.}~\bibnamefont {Vanderstraeten}},\
  }\bibfield  {title} {\bibinfo {title} {Direct sampling of projected
  entangled-pair states},\ }\href {https://doi.org/10.1103/PhysRevB.104.235141}
  {\bibfield  {journal} {\bibinfo  {journal} {Phys. Rev. B}\ }\textbf {\bibinfo
  {volume} {104}},\ \bibinfo {pages} {235141} (\bibinfo {year}
  {2021})}\BibitemShut {NoStop}%
\bibitem [{\citenamefont {Carleo}\ \emph {et~al.}(2012)\citenamefont {Carleo},
  \citenamefont {Becca}, \citenamefont {Schir{\'o}},\ and\ \citenamefont
  {Fabrizio}}]{tvmc1}%
  \BibitemOpen
  \bibfield  {author} {\bibinfo {author} {\bibfnamefont {G.}~\bibnamefont
  {Carleo}}, \bibinfo {author} {\bibfnamefont {F.}~\bibnamefont {Becca}},
  \bibinfo {author} {\bibfnamefont {M.}~\bibnamefont {Schir{\'o}}},\ and\
  \bibinfo {author} {\bibfnamefont {M.}~\bibnamefont {Fabrizio}},\ }\bibfield
  {title} {\bibinfo {title} {Localization and glassy dynamics of many-body
  quantum systems},\ }\href {https://doi.org/10.1038/srep00243} {\bibfield
  {journal} {\bibinfo  {journal} {Scientific Reports}\ }\textbf {\bibinfo
  {volume} {2}},\ \bibinfo {pages} {243} (\bibinfo {year} {2012})}\BibitemShut
  {NoStop}%
\bibitem [{\citenamefont {Carleo}\ \emph {et~al.}(2014)\citenamefont {Carleo},
  \citenamefont {Becca}, \citenamefont {Sanchez-Palencia}, \citenamefont
  {Sorella},\ and\ \citenamefont {Fabrizio}}]{tvmc2}%
  \BibitemOpen
  \bibfield  {author} {\bibinfo {author} {\bibfnamefont {G.}~\bibnamefont
  {Carleo}}, \bibinfo {author} {\bibfnamefont {F.}~\bibnamefont {Becca}},
  \bibinfo {author} {\bibfnamefont {L.}~\bibnamefont {Sanchez-Palencia}},
  \bibinfo {author} {\bibfnamefont {S.}~\bibnamefont {Sorella}},\ and\ \bibinfo
  {author} {\bibfnamefont {M.}~\bibnamefont {Fabrizio}},\ }\bibfield  {title}
  {\bibinfo {title} {Light-cone effect and supersonic correlations in one- and
  two-dimensional bosonic superfluids},\ }\href
  {https://doi.org/10.1103/PhysRevA.89.031602} {\bibfield  {journal} {\bibinfo
  {journal} {Phys. Rev. A}\ }\textbf {\bibinfo {volume} {89}},\ \bibinfo
  {pages} {031602} (\bibinfo {year} {2014})}\BibitemShut {NoStop}%
\bibitem [{\citenamefont {Ido}\ \emph {et~al.}(2015)\citenamefont {Ido},
  \citenamefont {Ohgoe},\ and\ \citenamefont {Imada}}]{tvmc3}%
  \BibitemOpen
  \bibfield  {author} {\bibinfo {author} {\bibfnamefont {K.}~\bibnamefont
  {Ido}}, \bibinfo {author} {\bibfnamefont {T.}~\bibnamefont {Ohgoe}},\ and\
  \bibinfo {author} {\bibfnamefont {M.}~\bibnamefont {Imada}},\ }\bibfield
  {title} {\bibinfo {title} {Time-dependent many-variable variational monte
  carlo method for nonequilibrium strongly correlated electron systems},\
  }\href {https://doi.org/10.1103/PhysRevB.92.245106} {\bibfield  {journal}
  {\bibinfo  {journal} {Phys. Rev. B}\ }\textbf {\bibinfo {volume} {92}},\
  \bibinfo {pages} {245106} (\bibinfo {year} {2015})}\BibitemShut {NoStop}%
\bibitem [{\citenamefont {Bhanot}\ and\ \citenamefont
  {Creutz}(1980)}]{Bhanot1980}%
  \BibitemOpen
  \bibfield  {author} {\bibinfo {author} {\bibfnamefont {G.}~\bibnamefont
  {Bhanot}}\ and\ \bibinfo {author} {\bibfnamefont {M.}~\bibnamefont
  {Creutz}},\ }\bibfield  {title} {\bibinfo {title} {Phase diagram of
  \text{Z(N)} and \text{U(1)} gauge theories in three dimensions},\ }\href
  {https://doi.org/10.1103/PhysRevD.21.2892} {\bibfield  {journal} {\bibinfo
  {journal} {Phys. Rev. D}\ }\textbf {\bibinfo {volume} {21}},\ \bibinfo
  {pages} {2892} (\bibinfo {year} {1980})}\BibitemShut {NoStop}%
\bibitem [{\citenamefont {Apte}\ \emph {et~al.}(2024)\citenamefont {Apte},
  \citenamefont {C\'ordova}, \citenamefont {Huang},\ and\ \citenamefont
  {Ashmore}}]{Apte2024deep}%
  \BibitemOpen
  \bibfield  {author} {\bibinfo {author} {\bibfnamefont {A.}~\bibnamefont
  {Apte}}, \bibinfo {author} {\bibfnamefont {C.}~\bibnamefont {C\'ordova}},
  \bibinfo {author} {\bibfnamefont {T.-C.}\ \bibnamefont {Huang}},\ and\
  \bibinfo {author} {\bibfnamefont {A.}~\bibnamefont {Ashmore}},\ }\bibfield
  {title} {\bibinfo {title} {Deep learning lattice gauge theories},\ }\href
  {https://doi.org/10.1103/PhysRevB.110.165133} {\bibfield  {journal} {\bibinfo
   {journal} {Phys. Rev. B}\ }\textbf {\bibinfo {volume} {110}},\ \bibinfo
  {pages} {165133} (\bibinfo {year} {2024})}\BibitemShut {NoStop}%
\bibitem [{\citenamefont {Grosse}\ \emph {et~al.}(1981)\citenamefont {Grosse},
  \citenamefont {Lang},\ and\ \citenamefont {Nicolai}}]{grosse1981equivalence}%
  \BibitemOpen
  \bibfield  {author} {\bibinfo {author} {\bibfnamefont {H.}~\bibnamefont
  {Grosse}}, \bibinfo {author} {\bibfnamefont {C.}~\bibnamefont {Lang}},\ and\
  \bibinfo {author} {\bibfnamefont {H.}~\bibnamefont {Nicolai}},\ }\bibfield
  {title} {\bibinfo {title} {Equivalence of the z4 and the z2 × z2 lattice
  gauge theories},\ }\href
  {https://doi.org/https://doi.org/10.1016/0370-2693(81)90370-1} {\bibfield
  {journal} {\bibinfo  {journal} {Physics Letters B}\ }\textbf {\bibinfo
  {volume} {98}},\ \bibinfo {pages} {69} (\bibinfo {year} {1981})}\BibitemShut
  {NoStop}%
\bibitem [{\citenamefont {Jalabert}\ and\ \citenamefont
  {Sachdev}(1991)}]{sachdev1991}%
  \BibitemOpen
  \bibfield  {author} {\bibinfo {author} {\bibfnamefont {R.~A.}\ \bibnamefont
  {Jalabert}}\ and\ \bibinfo {author} {\bibfnamefont {S.}~\bibnamefont
  {Sachdev}},\ }\bibfield  {title} {\bibinfo {title} {Spontaneous alignment of
  frustrated bonds in an anisotropic, three-dimensional ising model},\ }\href
  {https://doi.org/10.1103/PhysRevB.44.686} {\bibfield  {journal} {\bibinfo
  {journal} {Phys. Rev. B}\ }\textbf {\bibinfo {volume} {44}},\ \bibinfo
  {pages} {686} (\bibinfo {year} {1991})}\BibitemShut {NoStop}%
\bibitem [{\citenamefont {Senthil}\ and\ \citenamefont
  {Fisher}(2000)}]{senthil2000}%
  \BibitemOpen
  \bibfield  {author} {\bibinfo {author} {\bibfnamefont {T.}~\bibnamefont
  {Senthil}}\ and\ \bibinfo {author} {\bibfnamefont {M.~P.~A.}\ \bibnamefont
  {Fisher}},\ }\bibfield  {title} {\bibinfo {title} {${Z}_{2}$ gauge theory of
  electron fractionalization in strongly correlated systems},\ }\href
  {https://doi.org/10.1103/PhysRevB.62.7850} {\bibfield  {journal} {\bibinfo
  {journal} {Phys. Rev. B}\ }\textbf {\bibinfo {volume} {62}},\ \bibinfo
  {pages} {7850} (\bibinfo {year} {2000})}\BibitemShut {NoStop}%
\bibitem [{\citenamefont {Moessner}\ \emph {et~al.}(2001)\citenamefont
  {Moessner}, \citenamefont {Sondhi},\ and\ \citenamefont
  {Fradkin}}]{fradkin2001}%
  \BibitemOpen
  \bibfield  {author} {\bibinfo {author} {\bibfnamefont {R.}~\bibnamefont
  {Moessner}}, \bibinfo {author} {\bibfnamefont {S.~L.}\ \bibnamefont
  {Sondhi}},\ and\ \bibinfo {author} {\bibfnamefont {E.}~\bibnamefont
  {Fradkin}},\ }\bibfield  {title} {\bibinfo {title} {Short-ranged resonating
  valence bond physics, quantum dimer models, and ising gauge theories},\
  }\href {https://doi.org/10.1103/PhysRevB.65.024504} {\bibfield  {journal}
  {\bibinfo  {journal} {Phys. Rev. B}\ }\textbf {\bibinfo {volume} {65}},\
  \bibinfo {pages} {024504} (\bibinfo {year} {2001})}\BibitemShut {NoStop}%
\bibitem [{\citenamefont {Sachdev}(2018)}]{Sachdev2019}%
  \BibitemOpen
  \bibfield  {author} {\bibinfo {author} {\bibfnamefont {S.}~\bibnamefont
  {Sachdev}},\ }\bibfield  {title} {\bibinfo {title} {Topological order,
  emergent gauge fields, and fermi surface reconstruction},\ }\href
  {https://doi.org/10.1088/1361-6633/aae110} {\bibfield  {journal} {\bibinfo
  {journal} {Reports on Progress in Physics}\ }\textbf {\bibinfo {volume}
  {82}},\ \bibinfo {pages} {014001} (\bibinfo {year} {2018})}\BibitemShut
  {NoStop}%
\bibitem [{\citenamefont {Wenzel}\ \emph {et~al.}(2012)\citenamefont {Wenzel},
  \citenamefont {Coletta}, \citenamefont {Korshunov},\ and\ \citenamefont
  {Mila}}]{FFTFIM2012}%
  \BibitemOpen
  \bibfield  {author} {\bibinfo {author} {\bibfnamefont {S.}~\bibnamefont
  {Wenzel}}, \bibinfo {author} {\bibfnamefont {T.}~\bibnamefont {Coletta}},
  \bibinfo {author} {\bibfnamefont {S.~E.}\ \bibnamefont {Korshunov}},\ and\
  \bibinfo {author} {\bibfnamefont {F.}~\bibnamefont {Mila}},\ }\bibfield
  {title} {\bibinfo {title} {Evidence for columnar order in the fully
  frustrated transverse field ising model on the square lattice},\ }\href
  {https://doi.org/10.1103/PhysRevLett.109.187202} {\bibfield  {journal}
  {\bibinfo  {journal} {Phys. Rev. Lett.}\ }\textbf {\bibinfo {volume} {109}},\
  \bibinfo {pages} {187202} (\bibinfo {year} {2012})}\BibitemShut {NoStop}%
\bibitem [{\citenamefont {Gonz\'alez-Cuadra}\ \emph {et~al.}(2019)\citenamefont
  {Gonz\'alez-Cuadra}, \citenamefont {Dauphin}, \citenamefont {Grzybowski},
  \citenamefont {W\'ojcik}, \citenamefont {Lewenstein},\ and\ \citenamefont
  {Bermudez}}]{Z2BH_2019}%
  \BibitemOpen
  \bibfield  {author} {\bibinfo {author} {\bibfnamefont {D.}~\bibnamefont
  {Gonz\'alez-Cuadra}}, \bibinfo {author} {\bibfnamefont {A.}~\bibnamefont
  {Dauphin}}, \bibinfo {author} {\bibfnamefont {P.~R.}\ \bibnamefont
  {Grzybowski}}, \bibinfo {author} {\bibfnamefont {P.}~\bibnamefont
  {W\'ojcik}}, \bibinfo {author} {\bibfnamefont {M.}~\bibnamefont
  {Lewenstein}},\ and\ \bibinfo {author} {\bibfnamefont {A.}~\bibnamefont
  {Bermudez}},\ }\bibfield  {title} {\bibinfo {title} {Symmetry-breaking
  topological insulators in the ${{\Z}}_{2}$ bose-hubbard model},\ }\href
  {https://doi.org/10.1103/PhysRevB.99.045139} {\bibfield  {journal} {\bibinfo
  {journal} {Phys. Rev. B}\ }\textbf {\bibinfo {volume} {99}},\ \bibinfo
  {pages} {045139} (\bibinfo {year} {2019})}\BibitemShut {NoStop}%
\bibitem [{\citenamefont {Fradkin}(2013)}]{Fradkin2013}%
  \BibitemOpen
  \bibfield  {author} {\bibinfo {author} {\bibfnamefont {E.}~\bibnamefont
  {Fradkin}},\ }\href@noop {} {\emph {\bibinfo {title} {Fideld Theories of
  Condensed Matter Physics}}}\ (\bibinfo  {publisher} {Cambridge University
  Press},\ \bibinfo {year} {2013})\BibitemShut {NoStop}%
\bibitem [{\citenamefont {Carleo}\ \emph {et~al.}(2017)\citenamefont {Carleo},
  \citenamefont {Cevolani}, \citenamefont {Sanchez-Palencia},\ and\
  \citenamefont {Holzmann}}]{nqstvmc_1}%
  \BibitemOpen
  \bibfield  {author} {\bibinfo {author} {\bibfnamefont {G.}~\bibnamefont
  {Carleo}}, \bibinfo {author} {\bibfnamefont {L.}~\bibnamefont {Cevolani}},
  \bibinfo {author} {\bibfnamefont {L.}~\bibnamefont {Sanchez-Palencia}},\ and\
  \bibinfo {author} {\bibfnamefont {M.}~\bibnamefont {Holzmann}},\ }\bibfield
  {title} {\bibinfo {title} {Unitary dynamics of strongly interacting bose
  gases with the time-dependent variational monte carlo method in continuous
  space},\ }\href {https://doi.org/10.1103/PhysRevX.7.031026} {\bibfield
  {journal} {\bibinfo  {journal} {Phys. Rev. X}\ }\textbf {\bibinfo {volume}
  {7}},\ \bibinfo {pages} {031026} (\bibinfo {year} {2017})}\BibitemShut
  {NoStop}%
\bibitem [{\citenamefont {Schmitt}\ and\ \citenamefont
  {Heyl}(2020)}]{nqstvmc_3}%
  \BibitemOpen
  \bibfield  {author} {\bibinfo {author} {\bibfnamefont {M.}~\bibnamefont
  {Schmitt}}\ and\ \bibinfo {author} {\bibfnamefont {M.}~\bibnamefont {Heyl}},\
  }\bibfield  {title} {\bibinfo {title} {Quantum many-body dynamics in two
  dimensions with artificial neural networks},\ }\href
  {https://doi.org/10.1103/PhysRevLett.125.100503} {\bibfield  {journal}
  {\bibinfo  {journal} {Phys. Rev. Lett.}\ }\textbf {\bibinfo {volume} {125}},\
  \bibinfo {pages} {100503} (\bibinfo {year} {2020})}\BibitemShut {NoStop}%
\bibitem [{\citenamefont {Guti{\'{e}}rrez}\ and\ \citenamefont
  {Mendl}(2022)}]{nqstvmc_4}%
  \BibitemOpen
  \bibfield  {author} {\bibinfo {author} {\bibfnamefont {I.~L.}\ \bibnamefont
  {Guti{\'{e}}rrez}}\ and\ \bibinfo {author} {\bibfnamefont {C.~B.}\
  \bibnamefont {Mendl}},\ }\bibfield  {title} {\bibinfo {title} {Real time
  evolution with neural-network quantum states},\ }\href
  {https://doi.org/10.22331/q-2022-01-20-627} {\bibfield  {journal} {\bibinfo
  {journal} {{Quantum}}\ }\textbf {\bibinfo {volume} {6}},\ \bibinfo {pages}
  {627} (\bibinfo {year} {2022})}\BibitemShut {NoStop}%
\bibitem [{\citenamefont {Gartner}\ \emph {et~al.}(2022)\citenamefont
  {Gartner}, \citenamefont {Mazzanti},\ and\ \citenamefont
  {Zillich}}]{nqstvmc_5}%
  \BibitemOpen
  \bibfield  {author} {\bibinfo {author} {\bibfnamefont {M.}~\bibnamefont
  {Gartner}}, \bibinfo {author} {\bibfnamefont {F.}~\bibnamefont {Mazzanti}},\
  and\ \bibinfo {author} {\bibfnamefont {R.~E.}\ \bibnamefont {Zillich}},\
  }\bibfield  {title} {\bibinfo {title} {{Time-dependent variational Monte
  Carlo study of the dynamic response of bosons in an optical lattice}},\
  }\href {https://doi.org/10.21468/SciPostPhys.13.2.025} {\bibfield  {journal}
  {\bibinfo  {journal} {SciPost Phys.}\ }\textbf {\bibinfo {volume} {13}},\
  \bibinfo {pages} {025} (\bibinfo {year} {2022})}\BibitemShut {NoStop}%
\bibitem [{\citenamefont {Sinibaldi}\ \emph {et~al.}(2023)\citenamefont
  {Sinibaldi}, \citenamefont {Giuliani}, \citenamefont {Carleo},\ and\
  \citenamefont {Vicentini}}]{nqstvmc_6}%
  \BibitemOpen
  \bibfield  {author} {\bibinfo {author} {\bibfnamefont {A.}~\bibnamefont
  {Sinibaldi}}, \bibinfo {author} {\bibfnamefont {C.}~\bibnamefont {Giuliani}},
  \bibinfo {author} {\bibfnamefont {G.}~\bibnamefont {Carleo}},\ and\ \bibinfo
  {author} {\bibfnamefont {F.}~\bibnamefont {Vicentini}},\ }\bibfield  {title}
  {\bibinfo {title} {Unbiasing time-dependent {V}ariational {M}onte {C}arlo by
  projected quantum evolution},\ }\href
  {https://doi.org/10.22331/q-2023-10-10-1131} {\bibfield  {journal} {\bibinfo
  {journal} {{Quantum}}\ }\textbf {\bibinfo {volume} {7}},\ \bibinfo {pages}
  {1131} (\bibinfo {year} {2023})}\BibitemShut {NoStop}%
\bibitem [{\citenamefont {Medvidovi\ifmmode~\acute{c}\else \'{c}\fi{}}\ and\
  \citenamefont {Sels}(2023)}]{nqstvmc_7}%
  \BibitemOpen
  \bibfield  {author} {\bibinfo {author} {\bibfnamefont {M.}~\bibnamefont
  {Medvidovi\ifmmode~\acute{c}\else \'{c}\fi{}}}\ and\ \bibinfo {author}
  {\bibfnamefont {D.}~\bibnamefont {Sels}},\ }\bibfield  {title} {\bibinfo
  {title} {Variational quantum dynamics of two-dimensional rotor models},\
  }\href {https://doi.org/10.1103/PRXQuantum.4.040302} {\bibfield  {journal}
  {\bibinfo  {journal} {PRX Quantum}\ }\textbf {\bibinfo {volume} {4}},\
  \bibinfo {pages} {040302} (\bibinfo {year} {2023})}\BibitemShut {NoStop}%
\bibitem [{\citenamefont {Nys}\ \emph {et~al.}(2024)\citenamefont {Nys},
  \citenamefont {Pescia}, \citenamefont {Sinibaldi},\ and\ \citenamefont
  {Carleo}}]{nqstvmc_8}%
  \BibitemOpen
  \bibfield  {author} {\bibinfo {author} {\bibfnamefont {J.}~\bibnamefont
  {Nys}}, \bibinfo {author} {\bibfnamefont {G.}~\bibnamefont {Pescia}},
  \bibinfo {author} {\bibfnamefont {A.}~\bibnamefont {Sinibaldi}},\ and\
  \bibinfo {author} {\bibfnamefont {G.}~\bibnamefont {Carleo}},\ }\bibfield
  {title} {\bibinfo {title} {Ab-initio variational wave functions for the
  time-dependent many-electron schr{\"o}dinger equation},\ }\href
  {https://doi.org/10.1038/s41467-024-53672-w} {\bibfield  {journal} {\bibinfo
  {journal} {Nature Communications}\ }\textbf {\bibinfo {volume} {15}},\
  \bibinfo {pages} {9404} (\bibinfo {year} {2024})}\BibitemShut {NoStop}%
\bibitem [{Note1()}]{Note1}%
  \BibitemOpen
  \bibinfo {note} {We would like to thank Jutho Haegeman for sending us a
  recent unpublished master thesis from his group, where exact gradient descent
  using automatic differentiation was attempted for finding the ground states
  of an infinite-PEPS with structure similar to GCF. Although the numerical
  results are yet to converge with the bond dimension in the thesis, they do
  show the possibility of such calculations using more conventional methods
  based on double-layer TNS contraction.}\BibitemShut {Stop}%
\bibitem [{\citenamefont {Canals}\ \emph {et~al.}(2024)\citenamefont {Canals},
  \citenamefont {Chepiga},\ and\ \citenamefont {Tagliacozzo}}]{canals2024}%
  \BibitemOpen
  \bibfield  {author} {\bibinfo {author} {\bibfnamefont {M.}~\bibnamefont
  {Canals}}, \bibinfo {author} {\bibfnamefont {N.}~\bibnamefont {Chepiga}},\
  and\ \bibinfo {author} {\bibfnamefont {L.}~\bibnamefont {Tagliacozzo}},\
  }\bibfield  {title} {\bibinfo {title} {A tensor network formulation of
  lattice gauge theories based only on symmetric tensors},\ }\href
  {https://doi.org/10.48550/arXiv.2412.16961} {\bibfield  {journal} {\bibinfo
  {journal} {arXiv:2412.16961}\ } (\bibinfo {year} {2024})}\BibitemShut
  {NoStop}%
\bibitem [{\citenamefont {Hauschild}\ and\ \citenamefont
  {Pollmann}(2018)}]{hauschild2018efficient}%
  \BibitemOpen
  \bibfield  {author} {\bibinfo {author} {\bibfnamefont {J.}~\bibnamefont
  {Hauschild}}\ and\ \bibinfo {author} {\bibfnamefont {F.}~\bibnamefont
  {Pollmann}},\ }\bibfield  {title} {\bibinfo {title} {Efficient numerical
  simulations with tensor networks: Tensor network python (tenpy)},\
  }\href@noop {} {\bibfield  {journal} {\bibinfo  {journal} {SciPost Physics
  Lecture Notes}\ ,\ \bibinfo {pages} {005}} (\bibinfo {year}
  {2018})}\BibitemShut {NoStop}%
\bibitem [{\citenamefont {Hauschild}\ \emph {et~al.}(2024)\citenamefont
  {Hauschild}, \citenamefont {Unfried}, \citenamefont {Anand}, \citenamefont
  {Andrews}, \citenamefont {Bintz}, \citenamefont {Borla}, \citenamefont
  {Divic}, \citenamefont {Drescher}, \citenamefont {Geiger}, \citenamefont
  {Hefel}, \citenamefont {Hémery}, \citenamefont {Kadow}, \citenamefont
  {Kemp}, \citenamefont {Kirchner}, \citenamefont {Liu}, \citenamefont
  {Möller}, \citenamefont {Parker}, \citenamefont {Rader}, \citenamefont
  {Romen}, \citenamefont {Scalet}, \citenamefont {Schoonderwoerd},
  \citenamefont {Schulz}, \citenamefont {Soejima}, \citenamefont {Thoma},
  \citenamefont {Wu}, \citenamefont {Zechmann}, \citenamefont {Zweng},
  \citenamefont {Mong}, \citenamefont {Zaletel},\ and\ \citenamefont
  {Pollmann}}]{johannes2024tensor}%
  \BibitemOpen
  \bibfield  {author} {\bibinfo {author} {\bibfnamefont {J.}~\bibnamefont
  {Hauschild}}, \bibinfo {author} {\bibfnamefont {J.}~\bibnamefont {Unfried}},
  \bibinfo {author} {\bibfnamefont {S.}~\bibnamefont {Anand}}, \bibinfo
  {author} {\bibfnamefont {B.}~\bibnamefont {Andrews}}, \bibinfo {author}
  {\bibfnamefont {M.}~\bibnamefont {Bintz}}, \bibinfo {author} {\bibfnamefont
  {U.}~\bibnamefont {Borla}}, \bibinfo {author} {\bibfnamefont
  {S.}~\bibnamefont {Divic}}, \bibinfo {author} {\bibfnamefont
  {M.}~\bibnamefont {Drescher}}, \bibinfo {author} {\bibfnamefont
  {J.}~\bibnamefont {Geiger}}, \bibinfo {author} {\bibfnamefont
  {M.}~\bibnamefont {Hefel}}, \bibinfo {author} {\bibfnamefont
  {K.}~\bibnamefont {Hémery}}, \bibinfo {author} {\bibfnamefont
  {W.}~\bibnamefont {Kadow}}, \bibinfo {author} {\bibfnamefont
  {J.}~\bibnamefont {Kemp}}, \bibinfo {author} {\bibfnamefont {N.}~\bibnamefont
  {Kirchner}}, \bibinfo {author} {\bibfnamefont {V.~S.}\ \bibnamefont {Liu}},
  \bibinfo {author} {\bibfnamefont {G.}~\bibnamefont {Möller}}, \bibinfo
  {author} {\bibfnamefont {D.}~\bibnamefont {Parker}}, \bibinfo {author}
  {\bibfnamefont {M.}~\bibnamefont {Rader}}, \bibinfo {author} {\bibfnamefont
  {A.}~\bibnamefont {Romen}}, \bibinfo {author} {\bibfnamefont
  {S.}~\bibnamefont {Scalet}}, \bibinfo {author} {\bibfnamefont
  {L.}~\bibnamefont {Schoonderwoerd}}, \bibinfo {author} {\bibfnamefont
  {M.}~\bibnamefont {Schulz}}, \bibinfo {author} {\bibfnamefont
  {T.}~\bibnamefont {Soejima}}, \bibinfo {author} {\bibfnamefont
  {P.}~\bibnamefont {Thoma}}, \bibinfo {author} {\bibfnamefont
  {Y.}~\bibnamefont {Wu}}, \bibinfo {author} {\bibfnamefont {P.}~\bibnamefont
  {Zechmann}}, \bibinfo {author} {\bibfnamefont {L.}~\bibnamefont {Zweng}},
  \bibinfo {author} {\bibfnamefont {R.~S.~K.}\ \bibnamefont {Mong}}, \bibinfo
  {author} {\bibfnamefont {M.~P.}\ \bibnamefont {Zaletel}},\ and\ \bibinfo
  {author} {\bibfnamefont {F.}~\bibnamefont {Pollmann}},\ }\bibfield  {title}
  {\bibinfo {title} {{Tensor network Python (TeNPy) version 1}},\ }\href
  {https://doi.org/10.21468/SciPostPhysCodeb.41} {\bibfield  {journal}
  {\bibinfo  {journal} {SciPost Phys. Codebases}\ ,\ \bibinfo {pages} {41}}
  (\bibinfo {year} {2024})}\BibitemShut {NoStop}%
\bibitem [{\citenamefont {Vidal}(2003)}]{TEBD}%
  \BibitemOpen
  \bibfield  {author} {\bibinfo {author} {\bibfnamefont {G.}~\bibnamefont
  {Vidal}},\ }\bibfield  {title} {\bibinfo {title} {Efficient classical
  simulation of slightly entangled quantum computations},\ }\href
  {https://doi.org/10.1103/PhysRevLett.91.147902} {\bibfield  {journal}
  {\bibinfo  {journal} {Phys. Rev. Lett.}\ }\textbf {\bibinfo {volume} {91}},\
  \bibinfo {pages} {147902} (\bibinfo {year} {2003})}\BibitemShut {NoStop}%
\bibitem [{\citenamefont {Schwinger}(1962)}]{schwinger1962}%
  \BibitemOpen
  \bibfield  {author} {\bibinfo {author} {\bibfnamefont {J.}~\bibnamefont
  {Schwinger}},\ }\bibfield  {title} {\bibinfo {title} {Gauge invariance and
  mass. ii},\ }\href {https://doi.org/10.1103/PhysRev.128.2425} {\bibfield
  {journal} {\bibinfo  {journal} {Phys. Rev.}\ }\textbf {\bibinfo {volume}
  {128}},\ \bibinfo {pages} {2425} (\bibinfo {year} {1962})}\BibitemShut
  {NoStop}%
\bibitem [{\citenamefont {Shi}\ \emph {et~al.}(2009)\citenamefont {Shi},
  \citenamefont {Li}, \citenamefont {Zhao},\ and\ \citenamefont
  {Zhou}}]{shi2009graded}%
  \BibitemOpen
  \bibfield  {author} {\bibinfo {author} {\bibfnamefont {Q.-Q.}\ \bibnamefont
  {Shi}}, \bibinfo {author} {\bibfnamefont {S.-H.}\ \bibnamefont {Li}},
  \bibinfo {author} {\bibfnamefont {J.-H.}\ \bibnamefont {Zhao}},\ and\
  \bibinfo {author} {\bibfnamefont {H.-Q.}\ \bibnamefont {Zhou}},\ }\bibfield
  {title} {\bibinfo {title} {Graded projected entangled-pair state
  representations and an algorithm for translationally invariant strongly
  correlated electronic systems on infinite-size lattices in two spatial
  dimensions},\ }\href@noop {} {\bibfield  {journal} {\bibinfo  {journal}
  {arXiv preprint arXiv:0907.5520}\ } (\bibinfo {year} {2009})}\BibitemShut
  {NoStop}%
\bibitem [{\citenamefont {Corboz}\ \emph {et~al.}(2010)\citenamefont {Corboz},
  \citenamefont {Or\'us}, \citenamefont {Bauer},\ and\ \citenamefont
  {Vidal}}]{PhysRevB.81.165104}%
  \BibitemOpen
  \bibfield  {author} {\bibinfo {author} {\bibfnamefont {P.}~\bibnamefont
  {Corboz}}, \bibinfo {author} {\bibfnamefont {R.}~\bibnamefont {Or\'us}},
  \bibinfo {author} {\bibfnamefont {B.}~\bibnamefont {Bauer}},\ and\ \bibinfo
  {author} {\bibfnamefont {G.}~\bibnamefont {Vidal}},\ }\bibfield  {title}
  {\bibinfo {title} {Simulation of strongly correlated fermions in two spatial
  dimensions with fermionic projected entangled-pair states},\ }\href
  {https://doi.org/10.1103/PhysRevB.81.165104} {\bibfield  {journal} {\bibinfo
  {journal} {Phys. Rev. B}\ }\textbf {\bibinfo {volume} {81}},\ \bibinfo
  {pages} {165104} (\bibinfo {year} {2010})}\BibitemShut {NoStop}%
\bibitem [{Note2()}]{Note2}%
  \BibitemOpen
  \bibinfo {note} {Provided by Akira Matsumoto}\BibitemShut {NoStop}%
\bibitem [{\citenamefont {Dong}\ \emph {et~al.}(2019)\citenamefont {Dong},
  \citenamefont {Wang}, \citenamefont {Han}, \citenamefont {Guo},\ and\
  \citenamefont {He}}]{dong2019}%
  \BibitemOpen
  \bibfield  {author} {\bibinfo {author} {\bibfnamefont {S.-J.}\ \bibnamefont
  {Dong}}, \bibinfo {author} {\bibfnamefont {C.}~\bibnamefont {Wang}}, \bibinfo
  {author} {\bibfnamefont {Y.}~\bibnamefont {Han}}, \bibinfo {author}
  {\bibfnamefont {G.-c.}\ \bibnamefont {Guo}},\ and\ \bibinfo {author}
  {\bibfnamefont {L.}~\bibnamefont {He}},\ }\bibfield  {title} {\bibinfo
  {title} {Gradient optimization of fermionic projected entangled pair states
  on directed lattices},\ }\href {https://doi.org/10.1103/PhysRevB.99.195153}
  {\bibfield  {journal} {\bibinfo  {journal} {Phys. Rev. B}\ }\textbf {\bibinfo
  {volume} {99}},\ \bibinfo {pages} {195153} (\bibinfo {year}
  {2019})}\BibitemShut {NoStop}%
\bibitem [{\citenamefont {Dong}\ \emph {et~al.}(2020)\citenamefont {Dong},
  \citenamefont {Wang}, \citenamefont {Han}, \citenamefont {Yang},\ and\
  \citenamefont {He}}]{dong2020stable}%
  \BibitemOpen
  \bibfield  {author} {\bibinfo {author} {\bibfnamefont {S.-J.}\ \bibnamefont
  {Dong}}, \bibinfo {author} {\bibfnamefont {C.}~\bibnamefont {Wang}}, \bibinfo
  {author} {\bibfnamefont {Y.-J.}\ \bibnamefont {Han}}, \bibinfo {author}
  {\bibfnamefont {C.}~\bibnamefont {Yang}},\ and\ \bibinfo {author}
  {\bibfnamefont {L.}~\bibnamefont {He}},\ }\bibfield  {title} {\bibinfo
  {title} {Stable diagonal stripes in the $t$-${J}$ model at $\bar n_h=1/8$
  doping from \text{fPEPS} calculations},\ }\href
  {https://doi.org/10.1038/s41535-020-0226-4} {\bibfield  {journal} {\bibinfo
  {journal} {npj Quantum Materials}\ }\textbf {\bibinfo {volume} {5}},\
  \bibinfo {pages} {28} (\bibinfo {year} {2020})}\BibitemShut {NoStop}%
\bibitem [{\citenamefont {Becca}\ and\ \citenamefont
  {Sorella}(2017)}]{sorella2017}%
  \BibitemOpen
  \bibfield  {author} {\bibinfo {author} {\bibfnamefont {F.}~\bibnamefont
  {Becca}}\ and\ \bibinfo {author} {\bibfnamefont {S.}~\bibnamefont
  {Sorella}},\ }\href@noop {} {\emph {\bibinfo {title} {Quantum Monte Carlo
  Approaches for Correlated Systems}}}\ (\bibinfo  {publisher} {Cambridge
  University Press},\ \bibinfo {year} {2017})\BibitemShut {NoStop}%
\bibitem [{\citenamefont {Verstraete}\ \emph {et~al.}(2008)\citenamefont
  {Verstraete}, \citenamefont {Murg},\ and\ \citenamefont
  {Cirac}}]{verstraete2008}%
  \BibitemOpen
  \bibfield  {author} {\bibinfo {author} {\bibfnamefont {F.}~\bibnamefont
  {Verstraete}}, \bibinfo {author} {\bibfnamefont {V.}~\bibnamefont {Murg}},\
  and\ \bibinfo {author} {\bibfnamefont {J.~I.}\ \bibnamefont {Cirac}},\
  }\bibfield  {title} {\bibinfo {title} {Matrix product states, projected
  entangled pair states, and variational renormalization group methods for
  quantum spin systems},\ }\href {https://doi.org/10.1080/14789940801912366}
  {\bibfield  {journal} {\bibinfo  {journal} {Advances in Physics}\ }\textbf
  {\bibinfo {volume} {57}},\ \bibinfo {pages} {143} (\bibinfo {year}
  {2008})}\BibitemShut {NoStop}%
\bibitem [{\citenamefont {Vidal}(2007)}]{iTEBD}%
  \BibitemOpen
  \bibfield  {author} {\bibinfo {author} {\bibfnamefont {G.}~\bibnamefont
  {Vidal}},\ }\bibfield  {title} {\bibinfo {title} {Classical simulation of
  infinite-size quantum lattice systems in one spatial dimension},\ }\href
  {https://doi.org/10.1103/PhysRevLett.98.070201} {\bibfield  {journal}
  {\bibinfo  {journal} {Phys. Rev. Lett.}\ }\textbf {\bibinfo {volume} {98}},\
  \bibinfo {pages} {070201} (\bibinfo {year} {2007})}\BibitemShut {NoStop}%
\bibitem [{\citenamefont {Haegeman}\ \emph {et~al.}(2011)\citenamefont
  {Haegeman}, \citenamefont {Cirac}, \citenamefont {Osborne}, \citenamefont
  {Pi\ifmmode~\check{z}\else \v{z}\fi{}orn}, \citenamefont {Verschelde},\ and\
  \citenamefont {Verstraete}}]{TDVP}%
  \BibitemOpen
  \bibfield  {author} {\bibinfo {author} {\bibfnamefont {J.}~\bibnamefont
  {Haegeman}}, \bibinfo {author} {\bibfnamefont {J.~I.}\ \bibnamefont {Cirac}},
  \bibinfo {author} {\bibfnamefont {T.~J.}\ \bibnamefont {Osborne}}, \bibinfo
  {author} {\bibfnamefont {I.}~\bibnamefont {Pi\ifmmode~\check{z}\else
  \v{z}\fi{}orn}}, \bibinfo {author} {\bibfnamefont {H.}~\bibnamefont
  {Verschelde}},\ and\ \bibinfo {author} {\bibfnamefont {F.}~\bibnamefont
  {Verstraete}},\ }\bibfield  {title} {\bibinfo {title} {Time-dependent
  variational principle for quantum lattices},\ }\href
  {https://doi.org/10.1103/PhysRevLett.107.070601} {\bibfield  {journal}
  {\bibinfo  {journal} {Phys. Rev. Lett.}\ }\textbf {\bibinfo {volume} {107}},\
  \bibinfo {pages} {070601} (\bibinfo {year} {2011})}\BibitemShut {NoStop}%
\bibitem [{\citenamefont {Haegeman}\ \emph {et~al.}(2016)\citenamefont
  {Haegeman}, \citenamefont {Lubich}, \citenamefont {Oseledets}, \citenamefont
  {Vandereycken},\ and\ \citenamefont {Verstraete}}]{TDVP2}%
  \BibitemOpen
  \bibfield  {author} {\bibinfo {author} {\bibfnamefont {J.}~\bibnamefont
  {Haegeman}}, \bibinfo {author} {\bibfnamefont {C.}~\bibnamefont {Lubich}},
  \bibinfo {author} {\bibfnamefont {I.}~\bibnamefont {Oseledets}}, \bibinfo
  {author} {\bibfnamefont {B.}~\bibnamefont {Vandereycken}},\ and\ \bibinfo
  {author} {\bibfnamefont {F.}~\bibnamefont {Verstraete}},\ }\bibfield  {title}
  {\bibinfo {title} {Unifying time evolution and optimization with matrix
  product states},\ }\href {https://doi.org/10.1103/PhysRevB.94.165116}
  {\bibfield  {journal} {\bibinfo  {journal} {Phys. Rev. B}\ }\textbf {\bibinfo
  {volume} {94}},\ \bibinfo {pages} {165116} (\bibinfo {year}
  {2016})}\BibitemShut {NoStop}%
\bibitem [{\citenamefont {Wu}\ \emph {et~al.}(2012)\citenamefont {Wu},
  \citenamefont {Deng},\ and\ \citenamefont {Prokof'ev}}]{wu2012phase}%
  \BibitemOpen
  \bibfield  {author} {\bibinfo {author} {\bibfnamefont {F.}~\bibnamefont
  {Wu}}, \bibinfo {author} {\bibfnamefont {Y.}~\bibnamefont {Deng}},\ and\
  \bibinfo {author} {\bibfnamefont {N.}~\bibnamefont {Prokof'ev}},\ }\bibfield
  {title} {\bibinfo {title} {Phase diagram of the toric code model in a
  parallel magnetic field},\ }\href
  {https://doi.org/10.1103/PhysRevB.85.195104} {\bibfield  {journal} {\bibinfo
  {journal} {Phys. Rev. B}\ }\textbf {\bibinfo {volume} {85}},\ \bibinfo
  {pages} {195104} (\bibinfo {year} {2012})}\BibitemShut {NoStop}%
\end{thebibliography}%
